\documentclass[twocolumn,showpacs,prb]{revtex4}
\usepackage{amssymb}
\usepackage{graphicx}
\usepackage{dcolumn}
\usepackage{bm}
\usepackage{epsfig}
\usepackage{CJK}

\begin{document}

\begin{CJK*}{Bg5}{bsmi}

\title{Spin-charge conversion in multiterminal Aharonov-Casher ring coupled to precessing ferromagnets: A charge conserving Floquet-nonequilibrium Green function approach}

\author{Son-Hsien Chen (³¯ªQ½å)}
\email{d92222006@ntu.edu.tw}
\affiliation{Department of Physics, National Taiwan University, Taipei 10617, Taiwan}
\author{Chien-Liang Chen}
\affiliation{Department of Physics, National Taiwan University, Taipei 10617, Taiwan}
\author{Farzad Mahfouzi}
\affiliation{Department of Physics and Astronomy, University
of Delaware, Newark, DE 19716-2570, USA}
\author{Ching-Ray Chang (±i¼y·ç)}
\email{crchang@phys.ntu.edu.tw}
\affiliation{Department of Physics, National Taiwan University, Taipei 10617, Taiwan}

\begin{abstract}
We derive a non-perturbative solution to the Floquet-nonequilibrium
Green function (Floquet-NEGF) describing open quantum systems periodically driven by
an external field of arbitrary strength of frequency. By adopting the
reduced-zone scheme, we obtain expressions rendering conserved
charge currents for \emph{any given maximum number of photons},
distinguishable from other existed Floquet-NEGF-based expressions where, less feasible,
infinite number of photons needed to be taken into account to ensure the conservation.
To justify our derived formalism and to investigate spin-charge conversions by spin-orbit coupling (SOC), we consider the spin-driven setups as reciprocal
to the electric-driven setups in S. Souma \emph{et. al.}, Phys. Rev. B \textbf{70}, 195346
(2004) and Phys. Rev. Lett. \textbf{94}, 106602 (2005). In our setups,
\emph{pure} spin currents are driven by the magnetization dynamics of a
precessing ferromagnetic (FM) island and then are pumped into the
adjacent two- or four-terminal mesoscopic Aharonov-Casher (AC) ring of Rashba SOC where spin-charge conversions take place. Our spin-driven results show reciprocal features that excellently agree with the
findings in the electric-driven setups mentioned above.
We propose two types of symmetry operations,
under which the AC ring Hamiltonian is invariant,
to argue the relations of the pumped/converted currents in the leads within the same or between different pumping configurations. The symmetry arguments are independent of the ring width
and the number of open channels in the leads, terminals, and precessing FM islands,
In particular, \emph{net} pure in-plane spin currents and pure spin currents
can be generated in the leads for certain setups
of two terminals and two precessing FM islands with the
current magnitude and polarization direction
tunable by the pumping configuration, gate voltage covering the
two-terminal AC ring in between the FM islands.
\end{abstract}

\maketitle
\end{CJK*}

\section{Introduction}

\label{sec:intro} In this section, we first give the introduction to the
phenomenon and effects that motivate our investigation in Sec.~\ref%
{sec:intro_a}. An overview of the attempts of and findings in our study is
given in Sec.~\ref{sec:intro_b} where the organization of this paper is also
provided.

\subsection{Spin pumping, inverse spin-Hall effect, and Aharonov-Casher
effect without dc bias voltage}

\label{sec:intro_a} The spin-Hall effect (SHE) is a phenomenon where
longitudinal injection of a conventional unpolarized charge current into a
system with either extrinsic (due to impurities~\cite%
{D'yakonov1971b,Hirsch1999}) or intrinsic (due to band structure~\cite%
{Murakami2003a,Sinova2004}) spin-orbit coupling (SOC) generates a transverse
\emph{pure} spin current in the four-terminal geometry or the corresponding
spin accumulation along the lateral edges in the two-terminal geometry.
While the magnitude of the pure spin current generated by SHE in metals and
semiconductors is rather small and difficult to control,~\cite{Awschalom2007}
the inverse spin-Hall effect~\cite{Hirsch1999,Hankiewicz2005} (ISHE) has
recently emerged as the principal experimental tool to detect induction of
pure spin currents by different sources.

In the ISHE (which can be viewed~\cite{Hirsch1999,Hankiewicz2005} as the
Onsager reciprocal phenomenon of the direct SHE), a longitudinal spin
current generates a transverse charge current or voltage in an open circuit.
Experimental examples employing ISHE to detect pure spin current include: (%
\emph{i}) a pure spin current pumped by precessing magnetization of a single
ferromagnetic (FM) layer under ferromagnetic resonance (FMR) conditions with
detection by injecting the pumped current into an adjacent normal-metal
(NM), such as Pt, Pd, Au, and Mo, or semiconductor layer;~\cite%
{Saitoh2006,Mosendz2010} (\emph{ii}) spin currents generated in nonlocal
spin valves;~\cite{Valenzuela2006} (\emph{iii}) a transient ballistic pure
spin current injected~\cite{Werake2011} by a pair of laser pulses in GaAs
multiple quantum wells being converted into a charge current generated by
ISHE before the first electron-hole scattering event, thereby providing
unambiguous evidence for the intrinsic direct and inverse SHE.

The spin pumping~\cite{Tserkovnyak2005} by precessing magnetization is a
phenomenon where the moving magnetization of a single FM layer, driven by
microwave radiation under the FMR, emits spin current into adjacent NM
layers. The emitted spin current is pure in the sense that it is not
accompanied by any net charge flux. This effect is termed pumping because it
occurs in the absence of any dc bias voltage. Particularly, the detection of
pure spin currents pumped by magnetization dynamics has become a widely
employed technique~\cite{Mosendz2010} to characterize the effectiveness of
the charge-spin conversion by the SHE via measuring the material-specific
spin-Hall angle (i.e., the ratio of spin-Hall and charge conductivities).
The same ISHE-based technique is almost exclusively used in the very recent
observations of thermal spin pumping and magnon-phonon-mediated spin-Seebeck
effect.~\cite{Uchida2011} Also, spin pumping makes it possible to inject~%
\cite{Ando2011} spins into semiconductors with electrically tunable
efficiency across an Ohmic contact, evading the notorious problem~\cite%
{Rashba2000} of impedance mismatch between the FM conductor and
high-resistivity material.

On the theoretical side, the mechanisms for converting pumped pure spin
current into charge current, due to a region with intrinsic or extrinsic SOC
into which the pumped spin current is injected, have been analyzed in a
number of recent studies. For example, Ref.~\onlinecite{Ohe2008} has shown
that both transverse and longitudinal charge currents are generated in the
four-terminal Rashba-spin-split two-dimensional electron gases (2DEGs) of
square shape which is adjacent to the FM island with precessing
magnetization that pumps longitudinal pure spin current into the 2DEG. In
this scheme, the output charge current can be increased by increasing the
strength of the Rashba SOC in the 2DEG.

Furthermore, the recent alternative description~\cite{Takeuchi2010} of spin
pumping in FM$|$NM multilayers, which encompasses both the earlier
considered~\cite{Silsbee1979} nonlocal diffusion of the spin accumulation at
the FM$|$NM interface generated by magnetization precession and the
effective field described by the \textquotedblleft standard
model\textquotedblright ~\cite{Tserkovnyak2005} of spin pumping viewed as an
example of adiabatic quantum pumping that is captured by the Brouwer
scattering formula,~\cite{Brouwer1998} has shown that \emph{spin-charge
conversion} does not always occur and that the conversion depends
sensitively on the type of spin-orbit interactions. That is, unlike in FM$|$%
NM systems where spin-charge conversion is driven by the extrinsic SOC and
assumed to follow simple phenomenological prediction $\mathbf{j}_{c}\propto
\mathbf{S}\times \mathbf{j}_{s}$ ($\mathbf{j}_{c}$ is charge current
density, $\mathbf{S}$ is the spin polarization direction, and $\mathbf{j}%
_{s} $ is the injected spin current density), the pumped charge currents in
Rashba systems were found to deviate from this naive formula.

\begin{figure}[tbp]
\centerline{\psfig{file=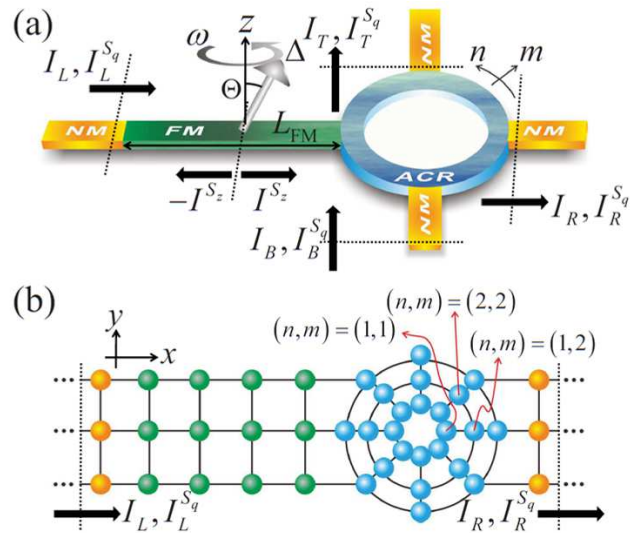,scale=1,angle=270,clip=true,trim=0mm 0mm
120mm 170mm}}
\caption{(Color online) Schematics of the (a) four- and (b) two-terminal
device setups. A precessing ferromagnetic (FM) island driven by microwaves
pumps pure spin current into the Aharonov-Casher (AC) ring (where the
spin-charge conversion takes place) patterned within a two-dimensional
electron gas in the $x$-$y$ plane with the Rashba spin-orbit coupling. The
semi-infinite normal-metal (NM) leads, in the absence of any dc bias
voltage, are attached to the AC ring to probe the output time-averaged (over
a precession period) pumped charge (spin-$q$ with $q\in \{x,y,z\}$)
currents, $I_{L}$ ($I_{L}^{S_{q}}$), $I_{B}$ ($I_{B}^{S_{q}}$), $I_{R}$ ($%
I_{R}^{S_{q}}$), and $I_{T}$ ($I_{T}^{S_{q}}$), in the left, bottom, right,
and top of the device, respectively. The FM island is of length $L_{FM}$,
and exchange splitting $\Delta $. The magnetization of the FM precesses
around the $z$-axis with frequency $\protect\omega $ and cone angle $\Theta $%
. The AC ring whose lattice sites along the tangential and normal directions
are denoted by $n$ and $m$, respectively.}
\label{fig1:setup}
\end{figure}

Thus, the whole phenomenon of spin-charge conversion after pure spin current
is injected into a system with SOC needs to be discussed together with the
origin of spin currents and the type of SOC employed for the conversion.~%
\cite{Takeuchi2010} Here we analyze spin current generation by one or two
precessing FM islands and the corresponding spin-charge conversion in two-
and four-terminal mesoscopic rings, adjacent to those islands and patterned
in the 2DEG with the Rashba SOC. Unlike the spin-charge conversion in
experimental and theoretical studies discussed above, where electronic
transport in semiclassical nature,~\cite{Takeuchi2010} the device depicted
in Fig.~\ref{fig1:setup} involves spin-sensitive quantum-interference
effects caused by the difference in the Aharonov-Casher (AC) phase~\cite%
{Aharonov1984,Mathur1992,Richter2012} gained by a spin traveling around the
\emph{phase-coherent} ring. The AC effect,~\cite%
{Aharonov1984,Mathur1992,Richter2012} in which magnetic dipoles travel
around a tube of electric charge, can be regarded as a special case of a
geometric phase. For typical ring sizes and strengths of Rashba SOC in
InAlAs/InGaAs heterostructures, the AC phase acquired by a (spin) magnetic
moment moving in the presence of an electrical field is of Aharonov-Anandan~%
\cite{Aharonov1987} (rather than Berry) type due to the fact that the
electron spin cannot~\cite{Richter2012} adiabatically maintain a fixed
orientation with respect to the radial effective (momentum-dependent and,
therefore, inhomogeneous) magnetic field associated with the Rashba SOC. In
fact, the AC phase for spins traveling around the mesoscopic ring consists
of not only the geometric phase, but also a dynamical phase arising from the
additional spin precession driven by the local effective magnetic field.~%
\cite{Frustaglia2004}

Accordingly, giving electrons such geometric phase makes it possible to
manipulate the magnitude of charge and spin currents in AC rings due to the
fact that, unlike usual case of intrinsically fixed phases, the ring
experiments allow one to steer geometric phases in a controlled way through
the system geometry and other various tunable parameters.~\cite{Nagasawa2012}
For example, the destructive quantum interferences, controlled by the
accumulated AC phase via tuning of the strength of the Rashba SOC (which
depends on the applied top-gate voltage~\cite{Nitta1997,Grundler2000}),
cause unpolarized charge current injected into the two-terminal AC rings to
diminish~\cite%
{Konig2006,Nitta2007,Nagasawa2012,Frustaglia2004a,Molnar2004,Souma2004} [to
zero~\cite{Souma2004} if the ring is strictly one-dimensional (1D)].
Similarly, in four-terminal AC rings one encounters
quantum-interference-controlled SHE, predicted in Ref.~%
\onlinecite{Souma2005a} and extended to different types of SOC and ring
geometry in Refs.~\onlinecite{Tserkovnyak2007} and~\onlinecite{Borunda2008},
where spin-Hall conductance can be tuned from zero to a finite value of the
order of spin conductance quantum $e/4\pi $.

\subsection{Methodology and key results}

\label{sec:intro_b} The goal of this study is threefold: (\emph{i}) to
provide a unified microscopic quantum transport theory based on the
non-perturbative solution of the time-dependent nonequilibrium Green
function (NEGF) in the Floquet representation~\cite{Tsuji2008} which \emph{%
conserves charge current at each level of approximation} (i.e., number of
microwave photons taken into account depending on the strength of the
driving field) for both the spin current generation by the magnetization
dynamics and spin-charge conversion in the adjacent region with SOC; (\emph{%
ii}) to understand how output spin and charge currents from multiterminal AC
ring device (such as in Fig.~\ref{fig1:setup}) can be controlled by the
top-gate covering the ring, by the cone angle of precessing magnetization
set by the input microwave power driving the precession, and by the setup
geometry; (\emph{iii}) to examine if the device setup in Fig.~\ref%
{fig1:setup} can be used as a new playground for experiments~\cite%
{Richter2012,Konig2006,Nitta2007,Nagasawa2012} measuring charge currents to
detect quantum interference effects involving AC phase in a single
mesoscopic ring where multichannel effects in a typical ring of finite width
act as effective dephasing (by entangling spin and orbital degrees of
freedom~\cite{Nikolic2005} or averaging over orbital channels with different
interference patterns~\cite{Souma2004}), thereby randomizing interference
patterns as in conventional measurements using dc bias voltage.~\cite%
{Konig2006}

The paper is organized as follows. In Sec.~\ref{sec:deha}, we specify our
pumping device and the adopted Hamiltonian. Section~\ref{sec:fgff}
formulates the solution to the Floquet-NEGF equations. Our numerical results
are discussed in Sec.~\ref{sec:rlt_dis} according to the chosen parameters
and units given in Sec.~\ref{sec:pa_un}. In Sec.~\ref{sec:ac_ishe}, we
examine both the pumped charge and spin currents responsible for the AC
phase and ISHE effects and driven by spin-pumping in the \emph{absence} of
any dc bias voltage, i.e., the spin-driven setups as the counterparts to the
conventional voltage-bias driven (electric-driven) setups with two-terminal~%
\cite{Frustaglia2004a,Molnar2004,Souma2004} and four-terminal~\cite%
{Souma2005a} mesoscopic AC rings of the Rashba SOC. Section~\ref{sec:sysinv}
illustrates different pumping symmetries of the AC ring. We conclude in Sec.~%
\ref{sec:conc}.

Our key results are as follows: (\emph{i}) To arrive at Eqs. (\ref{eq:I_pt}%
), (\ref{eq:I_pt^S_q}), (\ref{eq:I_p}), and (\ref{eq:I_p^S_q}), we solve the
Floquet-NEGF equations and use the so-called reduced-zone scheme~\cite%
{Tsuji2008} which guarantees conservation of charge currents for any given
maximum number of photons, unlike other recent approaches based on
continued-fraction solutions~\cite%
{Martinez2003,Kitagawa2011,Wang2003,Hattori2008} where charge conservation
is ensured only in the limit of infinite number of photons. (\emph{ii}) With
Fig.~\ref{fig2:2LLPM1} through Fig.~\ref{fig7:dis_4LLPM}, we analyze the
pumped currents in the spin-driven setup Fig. \ref{fig1:setup}. The results
are in good correspondences to the reciprocal electric-driven results shown
in Refs.~\onlinecite{Souma2004} and~\onlinecite{Souma2005a}, justifying the
derived formalism herein. Detail examinations, based on the AC effect and
ISHE, of the modulations of both the pumped charge and spin currents are
given. (\emph{iii}) In Sec.~\ref{sec:sysinv}, we tailor the pumping symmetry
under which the Hamiltonian of the AC ring of Rashba SOC remains invariant.
By performing the symmetry operations on one specific pumping configuration,
we can obtain the relations between pumped currents in the same or different
pumping configurations (or setup geometry). Although we illustrate the
symmetry operations by considering only the setups of two-terminal
two-precessing FM islands, the symmetry arguments are applicable to the case
of arbitrary number of terminals and FM islands as well, giving multifarious
manipulations of the pumped currents via setup geometry. In particular, Fig.~%
\ref{fig10:LPRP}, Fig.~\ref{fig11:LAPRAP}, and Fig.~\ref{fig14:H_LPRP}
through Fig.~\ref{fig17:H_LAPRP} show that the pumped spin currents are pure
and are of magnitude and polarization direction tunable by the top gate
voltage controlling the strength of the Rashba SOC and by the pumping
configurations.

\section{Device setup and Hamiltonian}

\label{sec:deha}

Consider the spin-driven four-terminal (or four-lead) setup in Fig.~\ref%
{fig1:setup}(a). A ferromagnet, FM, with precession axis along the $z$
direction contacts the AC ring of Rashba SOC in the $x$-$y$ plane from the
left. The FM plays the role of a spin-$z$ source, pumping pure spin-$z$
currents into the ring via the FM$|$AC-ring interface. The spin-charge
conversion takes place in the AC ring. The pumped or converted charge
current $I_{p}$ and spin current $I_{p}^{S_{q}}$ are probed by the NM leads
where currents are conserved with $p=$ $L$, $R $, $B$, and $T$ indicating
the currents flowing through the left, right, bottom, and top leads and $%
q\in \left\{ x,y,z\right\} $ standing for the pumped spin-$x$, $y$, and $z$
currents, respectively.

All computed pumped spin and charge currents are time-averaged (over one
precession period); they carry positive signs if the flow direction is in $%
+x $ or $+y$ direction or minus if flow direction is in $-x $ or $-y$. In
the two-terminal setup, Fig.~\ref{fig1:setup}(b), we have two NMs labeled by
$p=$ $L$, $R$. Note that, except the number of leads, Fig.~\ref{fig1:setup}%
(b) does not differ from Fig.~\ref{fig1:setup}(a), but just further shows
the lattice structure of the device. The AC ring is modeled by $m=1\cdots M$
concentric circles of the same number of lattice sites, and in a circle $m$
the lattice sites are indexed by $n=1\cdots N$. For instance, we have $%
(M,N)=(3,8)$ in Fig.~\ref{fig1:setup}(b). The NMs and the FM are modeled by
square lattices, while each NM is of semi-infinite length, and the FM is of
finite length, namely, an island.

The Hamiltonian of the whole device can be divided into six terms,%
\begin{eqnarray}
H\left( t\right) &=&H_{\text{ACR}}+H_{\text{NM}}+H_{\text{FM}}\left( t\right)
\nonumber \\
&&+H_{\text{NM-ACR}}+H_{\text{FM-ACR}}+H_{\text{FM-NM}}\text{,}
\label{eq:Hoft}
\end{eqnarray}%
where $H_{\text{ACR}}$, $H_{\text{NM}}$, and $H_{\text{FM}}$ account for the
Hamiltonian of the AC ring, NM, and FM, respectively. The term $H_{\text{%
NM-ACR}}$ describes the hybridization between NMs and AC ring, and $H_{\text{%
FM-ACR}}$ ($H_{\text{FM-NM}}$), the hybridization between FM and AC ring (FM
and NMs). Note that the time-dependent Hamiltonian originates only from the
precessing FM, $H_{\text{FM}}\left( t\right) $. Below, we express these six
terms explicitly.

Focus on $H_{\text{ACR}}$ first. As given in Ref.~\onlinecite{Souma2004},
the ring Hamiltonian can be written as,%
\begin{eqnarray}
H_{\text{ACR}} &=&\left[ \sum_{\sigma ,\sigma ^{\prime }=\uparrow
,\downarrow }\right. \varepsilon ^{n,m}\hat{a}_{n,m;\sigma }^{\dagger }\hat{a%
}_{n,m;\sigma }  \nonumber \\
&&-\left( \sum_{n=1}^{N}\sum_{m=1}^{M}\gamma _{\phi }^{n,n+1,m;\sigma
,\sigma ^{\prime }}\hat{a}_{n,m;\sigma }^{\dagger }\hat{a}_{n+1,m;\sigma
^{\prime }}\right.  \nonumber \\
&&\left. -\left. \sum_{n=1}^{N}\sum_{m=1}^{M-1}\gamma _{r}^{m,m+1,n;\sigma
,\sigma ^{\prime }}\hat{a}_{n,m;\sigma }^{\dagger }\hat{a}_{n,m+1;\sigma
^{\prime }}\right) \right]  \nonumber \\
&&+\text{H.c.,}  \label{eq:H_RR}
\end{eqnarray}%
with $n$ and $m$ denoting the lattice sites along the tangential ($\phi $)
and normal ($r$) directions as illustrated in Fig.~\ref{fig1:setup}(b),
respectively. The creation (annihilation) operator at site $(n,m)$ of spin $%
\sigma $ is $\hat{a}_{n,m;\sigma }^{\dagger }$ ($\hat{a}_{n,m;\sigma }$).
The on-site potential $\varepsilon ^{n,m}$ at site $(n,m)$ takes into
account the disorder and can be tuned by applying a top-gate voltage. In
what follows, unless further specified, we will assume that the AC ring, NM,
and FM, are all clean conductors, i.e., of zero on-site potentials. The
hopping along the $\phi $ direction,%
\begin{eqnarray}
\gamma _{\phi }^{n,n+1,m} &=&\frac{1}{\left( r_{m}/a\right) ^{2}\Delta \phi
^{2}}\gamma _{0}I_{s}  \nonumber \\
&&-i\frac{\gamma _{\text{SO}}}{\left( r_{m}/a\right) \Delta \phi }  \nonumber
\\
&&\times \left( \sigma _{x}\cos \phi _{n,n+1}+\sigma _{y}\sin \phi
_{n,n+1}\right) \text{,}  \label{eq:gamma_phi}
\end{eqnarray}%
and along the $r$ direction,
\begin{equation}
\gamma _{r}^{m,m+1,n}=\gamma _{0}I_{s}+i\gamma _{\text{SO}}\left( \sigma
_{y}\cos \phi _{n}-\sigma _{x}\sin \phi _{n}\right) \text{,}
\label{eq:gamma_r}
\end{equation}%
consists of two terms proportional to $\gamma _{0}$ that originates from the
kinetic energy and to $\gamma _{\text{SO}}$ that results from the Rashba
SOC, with $\phi _{n}\equiv 2\pi \left( n-1\right) /N$, $\phi _{n,n+1}\equiv
\left( \phi _{n}+\phi _{n+1}\right) /2$, $\Delta \phi =\phi _{2}-\phi _{1}$,
$r_{m}\equiv r_{1}+\left( m-1\right) a$, $\gamma _{0}=\hbar (2ma^{2})^{-1}$,
$\gamma _{\text{SO}}=\alpha \left( 2a\right) ^{-1},$ $a$ being the lattice
spacing, $\sigma _{q}=S_{q}2/\hbar $ being the Pauli matrices, and $\hbar
\times 2\pi $, the Planck constant. Being worth addressing, the Hamiltonian (%
\ref{eq:H_RR}) yields the same spin precession as obtained by the $SU$(2)
non-Abelian spin-orbit gauge~\cite{Chen2008} that absorbs the Rashba SOC
term for the U-shaped~\cite{Liu2011} 1D conductor; furthermore, the above
form of the concentric tight-binding Hamiltonian was also used to
theoretically model the Rashba SOC in HgTe/HgCdTe quantum wells in Ref.~%
\onlinecite{Konig2006}, showing experimental observations of the AC effect
in good agreements with the theoretical predictions, and thus strengthening
the validity of the ring Hamiltonian Eq. (\ref{eq:H_RR}).

The currents are probed by the un-biased NM leads whose Hamiltonian reads,
\begin{equation}
H_{\text{NM}}=-\sum_{p}\sum_{\sigma =\uparrow ,\downarrow
}\sum_{\left\langle \mu ,\mu ^{\prime }\right\rangle }\gamma _{0}\hat{b}%
_{\mu ;\sigma }^{\left( p\right) \dagger }\hat{b}_{\mu ^{\prime };\sigma
}^{\left( p\right) }\text{,}  \label{eq:H_NM}
\end{equation}%
where $\hat{b}_{\mu ;\sigma }^{\left( p\right) \dagger }$ is the creation
operator and $\hat{b}_{\mu ;\sigma }^{\left( p\right) }$ is the annihilation
operator in lead $p$ at site $\mu $ of spin $\sigma $. The pure spin
currents are pumped by the precessing FM described by,%
\begin{eqnarray}
H_{\text{FM}}\left( t\right) &=&\sum_{\sigma ,\sigma ^{\prime }=\uparrow
,\downarrow }\sum_{\nu }\frac{\Delta }{2}\vec{M}\left( t\right) \cdot \vec{%
\sigma}\hat{c}_{\nu ;\sigma }^{\dagger }\hat{c}_{\nu ;\sigma ^{\prime }}
\nonumber \\
&&-\sum_{\sigma =\uparrow ,\downarrow }\sum_{\left\langle \nu ,\nu ^{\prime
}\right\rangle }\gamma _{0}\hat{c}_{\nu ;\sigma }^{\dagger }\hat{c}_{\nu
^{\prime };\sigma }  \nonumber \\
&=&\sum_{\sigma ,\sigma ^{\prime }=\uparrow ,\downarrow }\sum_{\nu }\left[
Ve^{i\left( \omega t+\Phi \right) }+V^{\dagger }e^{-i\left( \omega t+\Phi
\right) }\right.  \nonumber \\
&&\left. +\frac{\Delta }{2}\cos \Theta \sigma _{z}\right] \hat{c}_{\nu
;\sigma }^{\dagger }\hat{c}_{\nu ^{\prime };\sigma ^{\prime }}  \nonumber \\
&&-\sum_{\sigma =\uparrow ,\downarrow }\sum_{\left\langle \nu ,\nu ^{\prime
}\right\rangle }\gamma _{0}\hat{c}_{\nu ;\sigma }^{\dagger }\hat{c}_{\nu
^{\prime };\sigma }\text{,}  \label{Eq:H_FM}
\end{eqnarray}%
with $\vec{M}\left( t\right) =\left[ \sin \Theta \cos \left( \omega t+\Phi
\right) ,\sin \theta \sin \left( \omega t+\Phi \right) ,\cos \Theta \right] $
giving $V=\sin \Theta \left( \sigma _{x}-i\sigma _{y}\right) \Delta /4$. The
$\hat{c}_{\nu ;\sigma }^{\dagger }$ ($\hat{c}_{\nu ;\sigma }$) is the
creation (annihilation) operator at site $\nu $ in the FM of spin $\sigma $.
The hybridizations between adjacent materials,%
\[
H_{\text{NM-ACR}}=-\gamma _{0}\sum_{p}\sum_{\sigma =\uparrow ,\downarrow
}\sum_{\left\langle \mu ,n\right\rangle }\hat{b}_{p;i;\sigma }^{\dagger }%
\hat{a}_{n;\sigma }+\text{H.c.,}
\]%
\[
H_{\text{FM-ACR}}=-\gamma _{0}\sum_{\sigma =\uparrow ,\downarrow
}\sum_{\left\langle \nu ,n\right\rangle }\hat{c}_{j;\sigma }^{\dagger }\hat{a%
}_{n;\sigma }+\text{H.c.,}
\]%
and%
\[
H_{\text{FM-NM}}=-\gamma _{0}\sum_{p}\sum_{\sigma =\uparrow ,\downarrow
}\sum_{\left\langle \nu ,\mu \right\rangle }\hat{c}_{j;\sigma }^{\dagger }%
\hat{b}_{j;i;\sigma }+\text{H.c.}
\]%
are set to be of the same strength, namely, $\gamma _{0}$.

\section{Floquet-nonequilibrium Green function approach for periodically
driven open quantum systems}

\label{sec:fgff}

In the devices where spin flip or spin precession is absent, the problem of
spin pumping by magnetization dynamics can be greatly simplified by mapping
it onto a time-independent one in the frame rotating with the precessing
magnetization.~\cite{Zhang2003a,Hattori2007,Tserkovnyak2008,Chen2009}
However, the device in Fig.~\ref{fig1:setup} contains Rashba SOC which
causes spin-up to evolve into to spin-down by spin precession, so that the
device Hamiltonian transformed in the rotating frame contains time-dependent
SOC terms.

In the adiabatic regime $\omega \rightarrow 0$, which is satisfied for
pumping by magnetization dynamics since the energy of microwave photons $%
\hbar\omega$ is much smaller than other relevant energy scales,~\cite%
{Mahfouzi2012} one can employ the Brouwer scattering formula.~\cite%
{Brouwer1998} However, this is numerically very inefficient since all pumped
spin and charge currents in devices, where the precessing FM island is
coupled to a region with SOC, are time-dependent.~\cite{Ohe2008} Thus, one
has to compute scattering matrix of the device repeatedly at each time step
of a discrete grid covering one period of harmonic external potential in
order to find full ac current vs. time dependence and then extract \emph{%
experimentally measured} dc component.

The relevant dc component of pumped current can be obtained from approaches
which generalize the existing steady-state transport theories, such as the
scattering matrix,~\cite{Moskalets2002,Wu2006b} NEGF formalism,~\cite%
{Tsuji2008,Martinez2003,Arrachea2005,FoaTorres2005,Wu2008,Wu2010,Kitagawa2011}%
and quantum master equations~\cite{Wu2010a} with the help of the Floquet
theorem~\cite{Floquet1883} valid for periodically driven systems. While the
equations of the Floquet-NEGF formalism we adopt here have been used before
to study a variety of charge pumping problems in non-interacting~\cite%
{Martinez2003,Arrachea2005,FoaTorres2005,Wu2008} and interacting electron
systems~\cite{Tsuji2008,Wu2010} or the photon-assisted dc transport,~\cite%
{Kitagawa2011} the key issue is to find a solution to these equations that
can capture pumping processes at arbitrary strength (or frequency) of the
external time-periodic potential while \emph{conserving}~\cite{Mahfouzi2012}
charge currents at each step of analytic or numerical algorithm. For
example, the often used continued fraction solution~\cite%
{Martinez2003,Kitagawa2011,Wang2003,Hattori2008} to Floquet-NEGF equations
does not~\cite{Mahfouzi2012} conserve charge current in the leads, and the
key trick we employ below to ensure current conservation is the \emph{%
reduced-zone scheme}.~\cite{Tsuji2008}

We begin the derivation for the charge-current-conserved Floquet-NEGF
solution by noting that the two fundamental objects~\cite{Haug2007} of the
NEGF formalism are the retarded
\begin{equation}
G_{\mathcal{I},\mathcal{J}}^{r}\left( t,t^{\prime }\right) \equiv -\frac{i}{%
\hbar }u\left( t-t^{\prime }\right) \left\langle \left\{ \hat{d}_{\mathcal{I}%
}\left( t\right) ,\hat{d}_{\mathcal{J}}^{\dagger }\left( t^{\prime }\right)
\right\} \right\rangle  \label{eq:def_Gr_ij}
\end{equation}%
and the lesser
\begin{equation}
G_{\mathcal{I},\mathcal{J}}^{<}\left( t,t^{\prime }\right) \equiv \frac{i}{%
\hbar }\left\langle \hat{d}_{\mathcal{J}}^{\dagger }\left( t^{\prime
}\right) \hat{d}_{\mathcal{I}}\left( t\right) \right\rangle \text{,}
\label{eq:def_G<_ij}
\end{equation}%
Green functions which describe the density of available quantum states and
how electrons occupy those states, respectively. Here $u$ is the unit step
function; indices $\{\mathcal{I},\mathcal{J}\}\in \{n,m,\mu ,\nu ,\sigma \}$%
and creation $\hat{d}^{\dagger }$ or annihilation operators $\hat{d}\in
\left\{ \hat{a},\hat{b},\hat{c}\right\}$ are used. For notational
convenience, the matrix representation with indices $\{\mathcal{I},\mathcal{J%
}\}$ will not be written out explicitly below.

The essence of the Floquet-NEGF approach is to treat the time variable $t$
in Eq. (\ref{eq:Hoft}) as an additional real-space degrees of freedom
denoted by $\check{t}$ with considering the auxiliary first-quantized
Hamiltonian
\begin{equation}
\check{h}\left( \check{t}\right) =h\left( \check{t}\right) -i\hbar \frac{%
\partial }{\partial \check{t}}\text{.}  \label{eq:defhc}
\end{equation}%
Here $h\left( \check{t}\right) $ is the first-quantization version of our
actual or original Hamiltonian (\ref{eq:Hoft}), i.e., the matrix
representation for $h\left( t\right) $ is of elements $h_{\mathcal{IJ}%
}\left( t\right) $ given by $H\left( t\right) =\sum_{\mathcal{I},\mathcal{J}}%
\hat{d}_{\mathcal{I}}^{\dagger }\hat{d}_{\mathcal{J}}h_{\mathcal{IJ}}\left(
t\right) $. The check-hatted symbol $\check{X}$ is used to remind us that $%
\check{X}$ is an auxiliary variable or operator but not the actual one.

The Schr\"{o}dinger equation for $\check{h}\left( \check{t}\right) $ reads,
\begin{equation}
i\hbar \frac{\partial }{\partial t}\check{\psi}\left( \check{t},t\right) =%
\check{h}\left( \check{t}\right) \check{\psi}\left( \check{t},t\right) \text{%
,}  \label{eq:EOM_phic}
\end{equation}%
while keeping in mind again that only $t$ is the real time variable, but $%
\check{t}$ is a virtual position variable. It is straightforward to prove
that, by assuming the wave function $\check{\psi}\left( \check{t},t\right) $
of the form $\check{\psi}\left( \check{t},t\right) =A\left( \check{t}\right)
B\left( t\right) $ in Eq.(\ref{eq:EOM_phic}) and then letting $\check{t}%
\rightarrow t$, the original wave function is recovered,
\begin{equation}
\check{\psi}\left( \check{t}=t,t\right) =\psi \left( t\right) \text{,}
\label{eq:phic}
\end{equation}%
where $\psi \left( t\right) $ obeys our original Schr\"{o}dinger equation $%
i\hbar \partial \psi \left( t\right) /\partial t=h\left( t\right) \psi
\left( t\right) $. Equation (\ref{eq:phic}) plays the fundamental role in
the Floquet-NEGF, since it bridges the two systems, the auxiliary
time-independent system described by $\check{h}\left( \check{t}\right) $ and
our original system described by $h\left( t\right) $. Accordingly, one can
first solve the problems in the time-independent system constructed
according to Eq. (\ref{eq:defhc}), express physical quantities or functions
in terms of $\check{\psi}\left( \check{t},t\right) $, and eventually set $%
\check{t}\rightarrow t$ to obtain the corresponding physical quantities or
functions for our original system.

To illustrate the idea above, consider the retarded Green function as an
example. The retarded Floquet Green function $\check{G}^{r}\left(
t,t^{\prime };\check{t},\check{t}^{\prime }\right) $ corresponding to our
auxiliary system $\check{h}\left( \check{t}\right) $ obeys the equation of
motion (EOM),%
\begin{equation}
\left[ i\hbar \frac{\partial }{\partial t}-\check{h}\left( \check{t}\right) %
\right] \check{G}^{r}\left( t,t^{\prime };\check{t},\check{t}^{\prime
}\right) =\delta \left( t-t^{\prime }\right) \delta _{T}\left( \check{t}-%
\check{t}^{\prime }\right)  \label{eq:Grc_EOM}
\end{equation}%
where $\delta _{T}\left( \check{t}-\check{t}^{\prime }\right) $ denotes the
Dirac delta function of period $T$. Note that $\check{h}\left( \check{t}%
\right) $ is time-independent; hence, $\check{G}^{r}\left( t,t^{\prime };%
\check{t},\check{t}^{\prime }\right) $ depends only on the single time
variable $t-t^{\prime }$ via the Fourier transformation%
\begin{equation}
\check{G}^{r}\left( t,t^{\prime };\check{t},\check{t}^{\prime }\right)
=\int_{-\infty }^{\infty }\frac{dE}{2\pi \hbar }e^{-iE\left( t-t^{\prime
}\right) /\hbar }\check{G}^{r}\left( E;\check{t},\check{t}^{\prime }\right)
\label{eq:Grc_FT}
\end{equation}%
and can be expanded by the wave functions of the form
\begin{widetext}
\begin{equation}
\check{G}^{r}\left( E;\check{t},\check{t}^{\prime }\right) =\sum_{n_{\text{ph%
}},m_{\text{ph}}=-\infty }^{\infty }\check{\psi}_{n_{\text{ph}}}\left(
\check{t}\right) \left[ \frac{1}{\left( E+i\eta \right) I-\check{h}}\right]
_{n_{\text{ph}},m_{\text{ph}}}\check{\psi}_{m_{\text{ph}}}^{\ast }\left(
\check{t}^{\prime }\right) \text{.}  \label{eq:Grc_waveexp}
\end{equation}%
\end{widetext}Here, the notations, identity operator $I$, $\eta \rightarrow
0^{+}$, $\left\{ n_{\text{ph}},m_{\text{ph}}\right\} \in $ integers, and $%
\left( \cdots \right) _{n_{\text{ph}},m_{\text{ph}}}\equiv
\int_{-T/2}^{T/2}dt\check{\psi}_{n_{\text{ph}}}^{\ast }\left( \check{t}%
\right) \left( \cdots \right) \check{\psi}_{m_{\text{ph}}}\left( \check{t}%
\right) $ are used, and the basis
\begin{equation}
\check{\psi}_{n_{\text{ph}}}\left( \check{t}\right) =\left( T\right)
^{-1/2}e^{-in_{\text{ph}}\omega \check{t}}  \label{eq:wavefun}
\end{equation}%
ensures the periodicity $\check{G}^{r}\left( E;\check{t}+lT,\check{t}%
^{\prime }+l^{\prime }T\right) $ with $\omega \equiv 2\pi /T$ and $\left\{
l,l^{\prime }\right\} \in $ integers. The $\left\{ \left[ \left( E+i\eta
\right) I-\check{h}\right] ^{-1}\right\} _{n_{\text{ph}},m_{\text{ph}}}$ in
Eq. (\ref{eq:Grc_waveexp}) is evaluated according to the definition (\ref%
{eq:defhc}) of $\check{h}$ via%
\begin{eqnarray}
\check{h}_{n_{\text{ph}},m_{\text{ph}}} &=&\int_{-T/2}^{T/2}d\check{t}\check{%
\psi}_{n_{\text{ph}}}^{\ast }\left( \check{t}\right) \check{h}\left( \check{t%
}\right) \check{\psi}_{m_{\text{ph}}}\left( \check{t}\right)  \nonumber \\
&=&h_{n_{\text{ph}},m_{\text{ph}}}-n_{\text{ph}}\hbar \omega \delta _{n_{%
\text{ph}},m_{\text{ph}}}\text{,}  \label{eq:hc_nm}
\end{eqnarray}%
with $n_{\text{ph}}<0$ ($n_{\text{ph}}>0$) accounting for the absorption
(emission) processes of photons, as indicated by the subscript
\textquotedblleft ph\textquotedblright , and
\begin{eqnarray}
h_{n_{\text{ph}},m_{\text{ph}}} &=&\int_{-T/2}^{T/2}d\check{t}\check{\psi}%
_{n_{\text{ph}}}^{\ast }\left( \check{t}\right) h\left( \check{t}\right)
\check{\psi}_{m_{\text{ph}}}\left( \check{t}\right)  \nonumber \\
&=&\frac{1}{T}\int_{-T/2}^{T/2}d\check{t}e^{i\left( n_{\text{ph}}-m_{\text{ph%
}}\right) \omega \check{t}}h\left( \check{t}\right) .  \label{eq:h_nm}
\end{eqnarray}%
Note that here beside the $\left\{ \mathcal{I},\mathcal{J}\right\} $ degrees
of freedoms, the extra degree of freedom, namely, photon is introduced. The $%
h_{n_{\text{ph}},m_{\text{ph}}}$ exists in the $\left\{ \mathcal{I},\mathcal{%
J}\right\} \otimes n_{\text{ph}}$ Hilbert space, and thus so does $\check{G}%
^{r}\left( E;\check{t},\check{t}^{\prime }\right) $. The primary result for
the actual (original) retarded Green function (\ref{eq:def_Gr_ij}) is
obtained by substituting (\ref{eq:Grc_waveexp}), computed by Eqs. (\ref%
{eq:wavefun}), (\ref{eq:hc_nm}), and (\ref{eq:h_nm}), into Eq. (\ref%
{eq:Grc_FT}), and then replacing $\check{t}$ with $t$ and $\check{t}^{\prime
}$ with $t^{\prime }$ in the wave-function-expanded Green function, namely $%
\left. \check{G}^{r}\left( t,t^{\prime };\check{t},\check{t}^{\prime
}\right) \right\vert _{\check{t}\rightarrow t,\check{t}^{\prime }\rightarrow
t^{\prime }}=G^{r}\left( t,t^{\prime }\right) $; nonetheless $G^{r}\left(
t,t^{\prime }\right) $ can be further simplified by noting the relation that
for $l_{\text{ph}}\in $ integers, one has%
\begin{widetext}
\begin{equation}
\left[ \frac{1}{\left( E+i\eta \right) I-\check{h}}\right] _{n_{\text{ph}%
},m_{\text{ph}}+l_{\text{ph}}}=\left[ \frac{1}{\left( E+l_{\text{ph}}\hbar
\omega +i\eta \right) -\check{h}}\right] _{n_{\text{ph}}-l_{\text{ph}},m_{%
\text{ph}}}\text{,}  \label{eq:shift}
\end{equation}
\end{widetext}which can be deduced simply from the observations that $%
\mathcal{M}\equiv $ $\left( E+i\eta \right) I-\check{h}$ is a matrix of
infinite size, and that $\left[ \left( E+i\eta \right) I-\check{h}\right]
_{n_{\text{ph}},m_{\text{ph}}+l_{\text{ph}}}$ and $\left[ \left( E+l_{\text{%
ph}}\hbar \omega +i\eta \right) -\check{h}\right] _{n_{\text{ph}}-l_{\text{ph%
}},m_{\text{ph}}}$ evaluated according to (\ref{eq:hc_nm}) and (\ref{eq:h_nm}%
) are at the same position of $\mathcal{M}$, i.e., the same matrix element
of $\mathcal{M}$; therefore, the element at the same matrix position of $%
\mathcal{M}^{-1}$ implies Eq. (\ref{eq:shift}), a manifestation of the
reduced-zone scheme in which the energy $E$ can be reduced to the zone of
range $\hbar \omega $. With the help of $\left\{ \left[ \left( E+i\eta
\right) I-\check{h}\right] ^{-1}\right\} _{n_{\text{ph}},m_{\text{ph}%
}}=\left\{ \left[ \left( E+m_{\text{ph}}\hbar \omega +i\eta \right) -\check{h%
}\right] ^{-1}\right\} _{n_{\text{ph}}-m_{\text{ph}},0}$ [$m_{\text{ph}}=0$
and the re-notation $l_{\text{ph}}\rightarrow m_{\text{ph}}$ in Eq. (\ref%
{eq:shift})] and change of variables, $E^{\prime }\equiv E+m_{\text{ph}%
}\hbar \omega $ and $n_{\text{ph}}^{\prime }\equiv n_{\text{ph}}-m_{\text{ph}%
}$, the retarded Green function now reads,
\begin{widetext}
\begin{eqnarray}
G^{r}\left( t,t^{\prime }\right)  &=&\left. \check{G}^{r}\left( t,t^{\prime
};\check{t},\check{t}^{\prime }\right) \right\vert _{\check{t}\rightarrow t,%
\check{t}^{\prime }\rightarrow t^{\prime }}  \nonumber \\
&=&\frac{1}{T}\int_{-\infty }^{\infty }\frac{dE}{2\pi \hbar }e^{-iE\left(
t-t^{\prime }\right) /\hbar }\sum_{n_{\text{ph}},m_{\text{ph}}=-\infty
}^{\infty }e^{-in_{\text{ph}}\omega t}e^{im_{\text{ph}}\omega t^{\prime }}%
\left[ \frac{1}{\left( E+i\eta \right) I-\check{h}}\right] _{n_{\text{ph}%
},m_{\text{ph}}}  \nonumber \\
&=&\delta \left( 0\right) \int_{-\infty }^{\infty }\frac{dE^{\prime }}{2\pi
\hbar }e^{-iE^{\prime }\left( t-t^{\prime }\right) /\hbar }\sum_{n_{\text{ph}%
}^{\prime }=-\infty }^{\infty }e^{-in_{\text{ph}}^{\prime }\omega t}\check{G}%
_{n_{\text{ph}}^{\prime },0}^{r}\left( E^{\prime }\right) \text{,}
\label{eq:Gr_tt'}
\end{eqnarray}%
\end{widetext}with $\check{G}_{n_{\text{ph}},m_{\text{ph}}}^{r}\left(
E\right) \equiv \left\{ \left[ \left( E+i\eta \right) I-\check{h}\right]
^{-1}\right\} _{n_{\text{ph}},m_{\text{ph}}}=\check{G}_{n_{\text{ph}}-m_{%
\text{ph}},0}^{r}\left( E+m_{\text{ph}}\hbar \omega \right) $; note that the
prefactor $\delta \left( 0\right) $ in (\ref{eq:Gr_tt'}) is due to the
cancelation of $m_{\text{ph}}\hbar \omega $ in the exponent, i.e., $\delta
\left( 0\right) =\left( T\right) ^{-1}\sum_{m_{\text{ph}}}1=\left( T\right)
^{-1}\sum_{m_{\text{ph}}}e^{-im_{\text{ph}}\hbar \omega \times 0}$, while
for physical quantities, this prefactor $\delta \left( 0\right) $ is
irrelevant; instead, it is the normalized [absence of $\delta \left(
0\right) $ in (\ref{eq:Gr_tt'})] Green function,
\begin{eqnarray}
\bar{G}^{r}\left( t,t^{\prime }\right) &\equiv &\int_{-\infty }^{\infty }%
\frac{dE}{2\pi \hbar }e^{-iE\left( t-t^{\prime }\right) /\hbar }  \nonumber
\\
&&\sum_{n_{\text{ph}}=-\infty }^{\infty }e^{-in_{\text{ph}}\omega t}\check{G}%
_{n_{\text{ph}},0}^{r}\left( E\right) \text{,}  \label{eq:Grbar_tt'}
\end{eqnarray}%
that renders physical observable. One can also verify that the expression (%
\ref{eq:Grbar_tt'}) satisfies the EOM,%
\begin{equation}
\left[ i\hbar \frac{\partial }{\partial t}-h\left( t\right) \right] \bar{G}%
^{r}\left( t,t^{\prime }\right) =\delta _{T}\left( t-t^{\prime }\right)
\text{,}  \label{eq:Grbar_EOM}
\end{equation}%
by applying $\int_{-T/2}^{T/2}dt^{\prime }$ to both sides of the above Eq. (%
\ref{eq:Grbar_EOM}). Comparing the EOMs (\ref{eq:Grc_EOM}) and (\ref%
{eq:Grbar_EOM}), one clearly sees that the evolution of $\bar{G}^{r}\left(
t,t^{\prime }\right) $ is governed by the actual system $h\left( t\right) $,
while $\check{G}^{r}\left( t,t^{\prime };\check{t},\check{t}^{\prime
}\right) $ is by the auxiliary system $\check{h}\left( \check{t}\right) $.

The lesser Green function can be obtained in the same manner. The auxiliary
lesser Green function obeys the Keldysh integral equation,
\begin{widetext}
\begin{equation}
\check{G}^{<}\left( t,t^{\prime };\check{t},\check{t}^{\prime }\right)
=\int_{-\infty }^{\infty }\int_{-\infty }^{\infty
}dt_{1}dt_{2}\int_{-T/2}^{T/2}\int_{-T/2}^{T/2}d\check{t}_{1}d\check{t}_{2}%
\check{G}^{r}\left( t,t_{1};\check{t},\check{t}_{1}\right) \check{\Sigma}%
^{<}\left( t_{1},t_{2};\check{t}_{1},\check{t}_{2}\right) \check{G}%
^{a}\left( t_{2},t^{\prime };\check{t}_{2},\check{t}^{\prime }\right) \text{.%
}  \label{eq:G<c_tt'tctc'}
\end{equation}
\end{widetext}The time-independent $\check{h}\left( \check{t}\right) $
allows $\check{G}^{<}\left( t,t^{\prime };\check{t},\check{t}^{\prime
}\right) $ to be expressed in terms of the single time variable $t-t^{\prime
}$ of the form,%
\begin{equation}
\check{G}^{<}\left( t,t^{\prime };\check{t},\check{t}^{\prime }\right)
=\int_{-\infty }^{\infty }\frac{dE}{2\pi \hbar }e^{-iE\left( t-t^{\prime
}\right) /\hbar }\check{G}^{<}\left( E;\check{t},\check{t}^{\prime }\right)
\text{,}  \label{eq:G<c_FT}
\end{equation}%
with $\check{G}^{<}\left( E;\check{t},\check{t}^{\prime }\right) $ being Eq.
(\ref{eq:G<c_tt'tctc'}) written in the energy domain,
\begin{widetext}
\begin{equation}
\check{G}^{<}\left( E;\check{t},\check{t}^{\prime }\right)
=\int_{-T/2}^{T/2}\int_{-T/2}^{T/2}d\check{t}_{1}d\check{t}_{2}\check{G}%
^{r}\left( E;\check{t},\check{t}_{1}\right) \check{\Sigma}^{<}\left( E;%
\check{t}_{1},\check{t}_{2}\right) \check{G}^{a}\left( E;\check{t}_{2},%
\check{t}^{\prime }\right) \text{.}  \label{eq:G<_Etctc'}
\end{equation}%
\end{widetext}Here $\check{G}^{a}\left( E;\check{t},\check{t}^{\prime
}\right) =\left[ \check{G}^{r}\left( E;\check{t}^{\prime },\check{t}\right) %
\right] ^{\dagger }$ is the advanced Green function, and $\check{\Sigma}%
^{<}\left( E;\check{t},\check{t}^{\prime }\right) =\left\vert \gamma
_{0}\right\vert ^{2}\check{g}^{<}\left( E;\check{t},\check{t}^{\prime
}\right) =\sum_{p}\left\vert \gamma _{0}\right\vert ^{2}\check{g}^{\left(
p\right) <}\left( E;\check{t},\check{t}^{\prime }\right) $ is the lesser
self energy accounting for the interactions from all probes with the bare
(probes that are free of interacting with the environments) lesser Green
function of probe $p$ denoted by $\check{g}^{\left( p\right) <}\left( E;%
\check{t},\check{t}^{\prime }\right) $. The primary result for the lesser
Green function (\ref{eq:def_G<_ij}) is obtained via $\left. \check{G}%
^{<}\left( t,t^{\prime };\check{t},\check{t}^{\prime }\right) \right\vert _{%
\check{t}\rightarrow t,\check{t}^{\prime }\rightarrow t^{\prime
}}=G^{<}\left( t,t^{\prime }\right) $ where $\check{G}^{<}\left( t,t^{\prime
};\check{t},\check{t}^{\prime }\right) $ is computed by the Green functions
in the \textit{time-independent} system $\check{h}\left( \check{t}\right) $
according to Eqs. (\ref{eq:G<c_FT}) and (\ref{eq:G<_Etctc'}). Similarly, $%
G^{<}\left( t,t^{\prime }\right) $ can also be further simplified by taking
advantages of relation (\ref{eq:shift}) and change of variables. For this
simplification, we utilize the Keldysh equation in the energy domain Eq. (%
\ref{eq:G<_Etctc'}) and the wave-function expansion to obtain the expression
of $\check{G}^{<}\left( t,t^{\prime };\check{t},\check{t}^{\prime }\right) $%
,
\begin{widetext}
\begin{eqnarray}
\check{G}^{<}\left( t,t^{\prime };\check{t},\check{t}^{\prime }\right)  &=&%
\frac{1}{T}\int_{-\infty }^{\infty }\frac{dE}{2\pi \hbar }e^{-iE\left(
t-t^{\prime }\right) /\hbar }\sum_{n_{\text{ph}},m_{\text{ph}}=-\infty
}^{\infty }\sum_{k_{\text{ph}},l_{\text{ph}}=-\infty }^{\infty }  \nonumber
\\
&&\times e^{-in_{\text{ph}}\omega \check{t}}\left[ \frac{1}{\left( E+i\eta
\right) I-\check{h}}\right] _{n_{\text{ph}},k_{\text{ph}}}\sum_{p}\left\vert
\gamma _{0}\right\vert ^{2}\check{g}_{k_{\text{ph}},l_{\text{ph}}}^{\left(
p\right) <}\left( E\right) \left[ \frac{1}{\left( E-i\eta \right) I-\check{h}%
^{\dagger }}\right] _{l_{\text{ph}},m_{\text{ph}}}e^{im_{\text{ph}}\omega
\check{t}^{\prime }}  \nonumber \\
&=&\frac{1}{T}\int_{-\infty }^{\infty }\frac{dE^{\prime }}{2\pi \hbar }%
e^{-i\left( E^{\prime }-m_{\text{ph}}\hbar \omega \right) \left( t-t^{\prime
}\right) /\hbar }  \nonumber \\
&&\times \sum_{n_{\text{ph}},m_{\text{ph}}=-\infty }^{\infty }e^{-in_{\text{%
ph}}\omega \check{t}}e^{im_{\text{ph}}\omega \check{t}^{\prime }}\sum_{k_{%
\text{ph}},l_{\text{ph}}^{\prime }=-\infty }^{\infty }\left[ \frac{1}{\left(
E^{\prime }-m_{\text{ph}}\hbar \omega +i\eta \right) I-\check{h}}\right]
_{n_{\text{ph}},k_{\text{ph}}}  \nonumber \\
&&\times \sum_{p}\left\vert \gamma _{0}\right\vert ^{2}\check{g}_{k_{\text{ph%
}},l_{\text{ph}}^{\prime }+m_{\text{ph}}}^{\left( p\right) <}\left(
E^{\prime }-m_{\text{ph}}\hbar \omega \right) \left[ \frac{1}{\left(
E^{\prime }-i\eta \right) I-\check{h}^{\dagger }}\right] _{l_{\text{ph}%
}^{\prime },0}\text{,}  \label{eq:G<c_step1}
\end{eqnarray}
\end{widetext}where $\left\{ \left[ \left( E-i\eta \right) I-\check{h}%
^{\dagger }\right] ^{-1}\right\} _{l_{\text{ph}},m_{\text{ph}}}=\left\{ %
\left[ \left( E+m_{\text{ph}}\hbar \omega -i\eta \right) I-\check{h}%
^{\dagger }\right] ^{-1}\right\} _{l_{\text{ph}}-m_{\text{ph}},0}$and change
of variables, $E^{\prime }\equiv E+m_{\text{ph}}\hbar \omega $ and $l_{\text{%
ph}}^{\prime }\equiv l_{\text{ph}}-m_{\text{ph}}$ are used. Note that the
\textit{original} Hamiltonian (\ref{eq:H_NM}) of the NM probes in our
present case is time-independent so that we have the expression, $\left\vert
\gamma _{0}\right\vert ^{2}\check{g}_{n_{\text{ph}},m_{\text{ph}}}^{\left(
p\right) <}\left( E\right) =-2i\check{f}_{n_{\text{ph}}}^{\left( p\right)
}\left( E\right) \left\vert \gamma _{0}\right\vert ^{2}\text{Im}\check{g}%
_{n_{\text{ph}},m_{\text{ph}}}^{\left( p\right) r}\left( E\right) =i\check{f}%
_{n_{\text{ph}}}^{\left( p\right) }\left( E\right) \check{\Gamma}_{n_{\text{%
ph}},m_{\text{ph}}}^{\left( p\right) }\left( E\right) $, with the
Fermi-Dirac distribution (of Fermi energy $E_{F}$ at zero temperature as the
regime we are interested in),%
\begin{eqnarray}
\check{f}_{n_{\text{ph}}}^{\left( p\right) }\left( E\right) &=&\check{f}%
^{\left( p\right) }\left( E+n_{\text{ph}}\hbar \omega \right)  \nonumber \\
&=&\lim_{\beta \rightarrow 0}\left[ 1+e^{\left( E+n_{\text{ph}}\hbar \omega
-E_{F}\right) /\beta }\right] ^{-1}  \label{eq:def_fd}
\end{eqnarray}%
and
\begin{widetext}
\begin{equation}
\check{\Gamma}_{n_{\text{ph}},m_{\text{ph}}}^{\left( p\right) }\left(
E\right) =i\left\vert \gamma _{0}\right\vert ^{2}\left\{ \left[ \left(
E+i\eta \right) I-\check{h}^{\left( p\right) }\right] ^{-1}-\left[ \left(
E-i\eta \right) I-\check{h}^{\left( p\right) \dagger }\right] ^{-1}\right\}
_{n_{\text{ph}},m_{\text{ph}}}\text{.}  \label{eq:def_Gamma}
\end{equation}%
\end{widetext}Here the definition (\ref{eq:defhc}) yields $\check{h}^{\left(
p\right) }\left( \check{t}\right) =h^{\left( p\right) }-i\hbar \partial
/\partial \check{t}$ with $h^{\left( p\right) }$ being the first-quantized
version of $H_{\text{NM}}=\sum_{p}h^{\left( p\right) }\hat{b}_{p}^{\dagger }%
\hat{b}_{p}$, and note again because $H_{\text{NM}}$ in Eq. (\ref{eq:H_NM})
or $h^{\left( p\right) }$ is time-independent, one has $\left[ h^{\left(
p\right) }\right] _{n_{\text{ph}},m_{\text{ph}}}=\delta _{n_{\text{ph}},m_{%
\text{ph}}}\left[ h^{\left( p\right) }\right] _{n_{\text{ph}}}$, resulting
in $\check{\Gamma}_{n_{\text{ph}},m_{\text{ph}}}^{\left( p\right) }\left(
E\right) $ proportional to $\delta _{n_{\text{ph}},m_{\text{ph}}}$, i.e.,
diagonal, and thus the bare lesser Green functions of NMs are diagonal in
photon (or Floquet) space $\check{g}_{n_{\text{ph}},m_{\text{ph}}}^{\left(
p\right) <}\left( E\right) =\delta _{n_{\text{ph}},m_{\text{ph}}}\check{g}%
_{n_{\text{ph}}}^{\left( p\right) <}\left( E\right) $ as well. Moreover,
applying the same argument we used to derive Eq. (\ref{eq:shift}) to $\check{%
g}_{n_{\text{ph}},m_{\text{ph}}}^{\left( p\right) <}\left( E\right) $
evaluated by Eqs. (\ref{eq:def_fd}) and (\ref{eq:def_Gamma}), one deduces,
\begin{equation}
\check{g}_{n_{\text{ph}},m_{\text{ph}}+l_{\text{ph}}}^{\left( p\right)
<}\left( E\right) =\check{g}_{n_{\text{ph}}-l_{\text{ph}},m_{\text{ph}%
}}^{\left( p\right) <}\left( E+l_{\text{ph}}\right) \text{,}
\label{eq:g<_shift}
\end{equation}%
which reflects again the reducible property (energy $E$ can be reduced to
the zone of range $\hbar \omega $) that yields the reduced-zone scheme.
Using above relation (\ref{eq:g<_shift}) and change of variables, $k_{\text{%
ph}}^{\prime }\equiv k_{\text{ph}}-m_{\text{ph}}$ and $n_{\text{ph}}^{\prime
}\equiv n_{\text{ph}}-m_{\text{ph}}$ in Eq. (\ref{eq:G<c_step1}), we arrive
at,%
\begin{widetext}
\begin{eqnarray}
\check{G}^{<}\left( t,t^{\prime };\check{t},\check{t}^{\prime }\right) &=&%
\frac{1}{T}\int_{-\infty }^{\infty }\frac{dE^{\prime }}{2\pi \hbar }%
e^{-i\left( E^{\prime }-m_{\text{ph}}\hbar \omega \right) \left( t-t^{\prime
}\right) /\hbar }\sum_{n_{\text{ph}}^{\prime },m_{\text{ph}}}e^{-i\left( n_{%
\text{ph}}^{\prime }+m_{\text{ph}}\right) \omega \check{t}}e^{im_{\text{ph}%
}\omega \check{t}^{\prime }}  \nonumber \\
&&\times \sum_{k_{\text{ph}}^{\prime },l_{\text{ph}}^{\prime }}\left[ \frac{1%
}{\left( E^{\prime }+i\eta \right) I-\check{h}}\right] _{n_{\text{ph}%
}^{\prime },k_{\text{ph}}^{\prime }}\sum_{p}\left\vert \gamma
_{0}\right\vert ^{2}\check{g}_{k_{\text{ph}}^{\prime },l_{\text{ph}}^{\prime
}}^{\left( p\right) <}\left( E^{\prime }\right) \left[ \frac{1}{\left(
E^{\prime }-i\eta \right) I-\check{h}^{\dagger }}\right] _{l_{\text{ph}%
}^{\prime },0}\text{,}  \label{eq:G<c_step2}
\end{eqnarray}%
\end{widetext}The actual lesser Green function can be obtained again via
setting $\check{t}\rightarrow t$ and $\check{t}^{\prime }\rightarrow
t^{\prime }$ in Eq. (\ref{eq:G<c_step2}) and noting that $m_{\text{ph}}\hbar
\omega $ in the exponent is canceled out; we thus have $\delta \left(
0\right) =\left( T\right) ^{-1}\sum_{m_{\text{ph}}}1=\left( T\right)
^{-1}\sum_{m_{\text{ph}}}e^{-im_{\text{ph}}\hbar \omega \times 0}$, so that
the actual lesser Green function can be written as,%
\begin{widetext}
\begin{eqnarray*}
G^{<}\left( t,t^{\prime }\right)  &=&\left. \check{G}^{<}\left( t,t^{\prime
};\check{t},\check{t}^{\prime }\right) \right\vert _{\check{t}\rightarrow t,%
\check{t}^{\prime }\rightarrow t^{\prime }} \\
&=&\delta \left( 0\right) \int_{-\infty }^{\infty }\frac{dE^{\prime }}{2\pi
\hbar }e^{-iE^{\prime }\left( t-t^{\prime }\right) /\hbar }\sum_{n_{\text{ph}%
}^{\prime }}e^{-in_{\text{ph}}^{\prime }\omega t}\sum_{k_{\text{ph}}^{\prime
},l_{\text{ph}}^{\prime }} \\
&&\times \left[ \frac{1}{\left( E^{\prime }+i\eta \right) I-\check{h}}\right]
_{n_{\text{ph}}^{\prime },k_{\text{ph}}^{\prime }}\sum_{p}\check{g}_{k_{%
\text{ph}}^{\prime },l_{\text{ph}}^{\prime }}^{\left( p\right) <}\left(
E^{\prime }\right) \left[ \frac{1}{\left( E^{\prime }-i\eta \right) I-\check{%
h}^{\dagger }}\right] _{l_{\text{ph}}^{\prime },0}\text{,}
\end{eqnarray*}%
and eventually obtain the normalized lesser Green function,%
\begin{eqnarray}
\bar{G}^{<}\left( t,t^{\prime }\right)  &=&\int_{-\infty }^{\infty }\frac{dE%
}{2\pi \hbar }e^{-iE\left( t-t^{\prime }\right) /\hbar }\sum_{n_{\text{ph}%
}}e^{-in_{\text{ph}}\omega t}\sum_{k_{\text{ph}},l_{\text{ph}}}  \nonumber \\
&&\times \left[ \frac{1}{\left( E+i\eta \right) I-\check{h}}\right] _{n_{%
\text{ph}},k_{\text{ph}}}\sum_{p}\left\vert \gamma _{0}\right\vert ^{2}%
\check{g}_{k_{\text{ph}},l_{\text{ph}}}^{\left( p\right) <}\left( E\right) %
\left[ \frac{1}{\left( E-i\eta \right) I-\check{h}^{\dagger }}\right] _{l_{%
\text{ph}},0}\text{.}  \label{eq:G<bar_tt'}
\end{eqnarray}
\end{widetext}

Physical quantities can be extracted from the actual Green function (\ref%
{eq:G<bar_tt'}). For instance, the quantum-statistical-averaged occupation
number at time $t$ on site $i\in \{n,m,\mu \}$ can be expressed as Tr$_{%
\text{s}}\left[ \bar{G}_{i,i}^{<}\left( t,t\right) O\right] \hbar /i$, while
the bond current from site $i$ to site $j$ reads,%
\begin{eqnarray}
J_{i\rightarrow j}\left( t\right) &=&-\text{Tr}_{\text{s}}\left[ \frac{%
\left\{ h_{j,i},O\right\} }{2}\bar{G}_{i,j}^{<}\left( t,t\right) \right.
\nonumber \\
&&\left. -\frac{\left\{ h_{i,j},O\right\} }{2}\bar{G}_{j,i}^{<}\left(
t,t\right) \right] \text{,}  \label{eq:J_ij(t)}
\end{eqnarray}%
with $O=I_{s}$ ($O=S_{q}$) for particle (spin $S_{q}$) occupations or
currents and $I_{s}$ being the $2\times 2$ identity matrix in the Pauli
space; the notation Tr$_{\text{s}}$ stands for performing the trace in the
Pauli space (spin Hilbert space), and the anticommutator $\{A,B\}$ is
defined as $AB+BA$. The particle (charge) current
\begin{widetext}
\begin{eqnarray}
I_{p}\left( t\right)  &=&\frac{1}{2\pi \hbar }\sum_{p^{\prime }}\sum_{n_{%
\text{ph}},m_{\text{ph}}}\int_{E_{F}-\hbar \omega /2}^{E_{F}+\hbar \omega
/2}dE\text{Tr}^{\prime }\left[ \check{G}^{r}\left( E\right) \check{f}%
^{\left( p^{\prime }\right) }\left( E\right) \check{\Gamma}^{\left(
p^{\prime }\right) }\left( E\right) \check{G}^{a}\left( E\right) \check{%
\Gamma}^{\left( p\right) }\left( E\right) \right.   \nonumber \\
&&-\left. \check{G}^{r}\left( E\right) \check{\Gamma}^{\left( p^{\prime
}\right) }\left( E\right) \check{G}^{a}\left( E\right) \check{f}^{\left(
p\right) }\left( E\right) \check{\Gamma}^{\left( p\right) }\left( E\right) %
\right] _{n_{\text{ph}},m_{\text{ph}}}e^{-i\left( n_{\text{ph}}-m_{\text{ph}%
}\right) \omega t}  \label{eq:I_pt}
\end{eqnarray}%
and spin current
\begin{eqnarray}
I_{p}^{S_{q}}\left( t\right)  &=&\frac{1}{4\pi }\sum_{p^{\prime }}\sum_{n_{%
\text{ph}},m_{\text{ph}}}\int_{E_{F}-\hbar \omega /2}^{E_{F}+\hbar \omega
/2}dE\text{Tr}^{\prime }\left[ \sigma _{q}\check{G}^{r}\left( E\right)
\check{f}^{\left( p^{\prime }\right) }\left( E\right) \check{\Gamma}^{\left(
p^{\prime }\right) }\left( E\right) \check{G}^{a}\left( E\right) \check{%
\Gamma}^{\left( p\right) }\left( E\right) \right.   \nonumber \\
&&-\left. \check{G}^{r}\left( E\right) \check{\Gamma}^{\left( p^{\prime
}\right) }\left( E\right) \check{G}^{a}\left( E\right) \check{f}^{\left(
p\right) }\left( E\right) \check{\Gamma}^{\left( p\right) }\left( E\right) %
\right] _{n_{\text{ph}},m_{\text{ph}}}e^{-i\left( n_{\text{ph}}-m_{\text{ph}%
}\right) \omega t}  \label{eq:I_pt^S_q}
\end{eqnarray}
\end{widetext}probed by lead $p$ are obtained by summing all the bond
currents (\ref{eq:J_ij(t)}) flowing through lead $p$. Note here the notation
Tr$^{\prime }$ performs the trace over all degrees of freedom, \emph{except}
photon's. All the functions within the trace are matrices; for example, the
Fermi-Dirac distribution is a matrix, $\check{f}^{\left( p\right) }\left(
E\right) =\check{f}_{n_{\text{ph}}}^{\left( p\right) }\left( E\right) \delta
_{n_{\text{ph}}\text{,}m_{\text{ph}}}$, computed by (\ref{eq:def_fd}). The
time-averaged currents can then be obtained via $I_{p}\equiv $sign$\left(
p\right) \times 2\pi \hbar T^{-1}\int_{-T/2}^{T/2}dtI_{p}\left( t\right) $
and $I_{p}^{S_{q}}\equiv $sign$\left( p\right) \times 4\pi
T^{-1}\int_{-T/2}^{T/2}dtI_{p}^{S_{q}}\left( t\right) $ as
\begin{widetext}
\begin{eqnarray}
I_{p} &=&\text{sign}\left( p\right) \times \sum_{p^{\prime
}}\int_{E_{F}-\hbar \omega /2}^{E_{F}+\hbar \omega /2}dE\text{Tr}\left[
\check{G}^{r}\left( E\right) \check{f}^{\left( p^{\prime }\right) }\left(
E\right) \check{\Gamma}^{\left( p^{\prime }\right) }\left( E\right) \check{G}%
^{a}\left( E\right) \check{\Gamma}^{\left( p\right) }\left( E\right) \right.
\nonumber \\
&&-\left. \check{G}^{r}\left( E\right) \check{\Gamma}^{\left( p^{\prime
}\right) }\left( E\right) \check{G}^{a}\left( E\right) \check{f}^{\left(
p\right) }\left( E\right) \check{\Gamma}^{\left( p\right) }\left( E\right) %
\right]   \label{eq:I_p}
\end{eqnarray}%
and%
\begin{eqnarray}
I_{p}^{S_{q}} &=&\text{sign}\left( p\right) \sum_{p^{\prime
}}\int_{E_{F}-\hbar \omega /2}^{E_{F}+\hbar \omega /2}dE\text{Tr}\left\{
\sigma _{q}\left[ \check{G}^{r}\left( E\right) \check{f}^{\left( p^{\prime
}\right) }\left( E\right) \check{\Gamma}^{\left( p^{\prime }\right) }\left(
E\right) \check{G}^{a}\left( E\right) \check{\Gamma}^{\left( p\right)
}\left( E\right) \right. \right.   \nonumber \\
&&-\left. \left. \check{G}^{r}\left( E\right) \check{\Gamma}^{\left(
p^{\prime }\right) }\left( E\right) \check{G}^{a}\left( E\right) \check{f}%
^{\left( p\right) }\left( E\right) \check{\Gamma}^{\left( p\right) }\left(
E\right) \right] \right\} \text{,}  \label{eq:I_p^S_q}
\end{eqnarray}
\end{widetext}with sign$\left( p\right) =1$ for $p\in \left\{ R,T\right\} $
and sign$\left( p\right) =-1$ for $p\in \left\{ L,B\right\} $. Notice here
the appearance of sign$\left( p\right) $ is merely for the sign convenience,
positive for right- or up-flowing currents, while negative for left- or
down-flowing currents. The prefactor $2\pi \hbar $ ($4\pi $) for $I_{p}$ ($%
I_{p}^{S_{q}}$) are adopted so that the units of $I_{p}$ and $I_{p}^{S_{q}}$
are the same. In other words, if $I_{p}$ measures the number of charge
quanta flowing through the lead $p$ per second, then $I_{p}^{S_{q}}$
measures the number of spins $S_{q}$ quanta flowing through the lead $p$ per
second. The trace here now is taken over all degrees of freedom, including
photon's.

We emphasize that, in Eqs. (\ref{eq:I_pt}), (\ref{eq:I_pt^S_q}), (\ref%
{eq:I_p}), and (\ref{eq:I_p^S_q}), the reduced-zone scheme~\cite{Tsuji2008}
such as (\ref{eq:shift}) or (\ref{eq:g<_shift}) is adopted, so that the
original integral interval $\left[ -\infty ,\infty \right] $ over energy $E$
is reduced to $\left[ E_{F}-\hbar \omega /2,E_{F}+\hbar \omega /2\right] $,
and with this scheme, for any given maximum $n_{\text{ph}}$, charge currents
are conserved, namely, $\sum_{p}I_{p}\left( t\right) =0$ or $\sum_{p}$sign$%
\left( p\right) I_{p}=0$. In principle all integers $n_{\text{ph}}$ should
account for transport, i.e., transitions involving any number of photons
have to be taken into account, nonetheless when the strength of the
time-dependent field is small, only few photons can be absorbed or emitted
by electrons near Fermi level, and thus considering transitions between
channels of few photons are sufficient enough to get accurate results of
currents. In our following calculations, $\left\vert n_{\text{ph}%
}\right\vert \leq 2$ is chosen, since we find $\left\vert n_{\text{ph}%
}\right\vert \leq 2$ and $\left\vert n_{\text{ph}}\right\vert \leq 3$ do not
yield significantly discernible results.

\section{Results and discussion}

\label{sec:rlt_dis}

By Eqs.~(\ref{eq:I_p}) and (\ref{eq:I_p^S_q}), we show and examine our
numerical results for two- and four-terminal spin-pumping setups with the
parameters and units specified in Sec.~\ref{sec:pa_un}. In Sec.~\ref%
{sec:ac_ishe}, we first concentrate on the two-terminal case [Fig. \ref%
{fig1:setup}(b)] to see the counterpart physics shown in Ref.~%
\onlinecite{Souma2004} and then, in Sec. \ref{sec:ishe}, we investigate the
four-terminal case [Fig. \ref{fig1:setup}(a)] to unveil the phenomena dual
to what were found in Ref.~\onlinecite{Souma2005a}. Our discussions are
restricted to the case of single precessing FM in Secs. \ref{sec:ac_ishe}
and \ref{sec:ishe}. In Sec.~\ref{sec:sysinv}, we aim at building up the
relations between probed currents in the same or different pumping
configurations from symmetry perspective; the two presented symmetries yield
invariant AC ring Hamiltonian, and the arguments on the relations based on
the symmetries are generally capable of setups of arbitrary number of
precessing FM islands and terminals. Nevertheless, for demonstration
simplicity, below in Sec. \ref{sec:sys_nresults}, where our numerical
results are shown to be in line with the predictions given by the symmetry
arguments, we consider only the two-terminal two-precessing-FM setups.

\subsection{Parameters and units}

\label{sec:pa_un}

The following parameters and units are used. All energies are in unit of the
hopping energy $\gamma _{0}$, and lengths are in unit of the lattice
constant $a$. For brevity, the aspect ratio $1/2$ between the length of FM, $%
L_{FM}$, and the length of AC ring, $N$, is adopted; for example, in Fig. %
\ref{fig1:setup}(b), we have $L_{FM}/N=4/8$. Also, the width of FM is set to
be the same as the width of NM. The default values of parameters of the
precession FM (FMs), the spin splitting strength $\Delta =1$, precession
frequency (energy) $\hbar \omega =10^{-3}$, precession cone angle $\Theta
=10^{\circ }$, and initial precession phase (azimuthal angle) $\Phi
=0^{\circ }$ are chosen. To compare the results of our spin-driven setups
with the findings of the electric-driven setups in Refs. %
\onlinecite{Souma2004} and \onlinecite{Souma2005a}, the size of the AC ring
are set similar or according to Refs. \onlinecite{Souma2004} and %
\onlinecite{Souma2005a}. We refer to the ring of $M=1$ as the strict 1D
ring, and $M>1$ as the quasi 1D ring. Note again that the sign convention
used here is, positive for right- or up-moving flow and negative for left-
or down-moving flow. No bias is applied to any probes for what we consider
here are all spin-driven setups. The number of open channels $M_{\text{open}%
} $ in the leads is adjustable by varying $E_{F}$; referring to Fig. 2 in
Ref. \onlinecite{Souma2004}, for leads of width consisting of three lattice
sites, one has $M_{\text{open}}=1$ approximately in the interval $E_{F}\in
\lbrack \pm 3.9,\pm 2]$, $M_{\text{open}}=2$ in $E_{F}\in \lbrack \pm 2,\pm
0.5]$, and $M_{\text{open}}=3$ in $E_{F}\in \lbrack -0.5,0.5]$.

\subsection{Single precessing FM island attached to two-terminal mesoscopic
AC ring}

\label{sec:ac_ishe}

\begin{figure}[tbp]
\centerline{\psfig{file=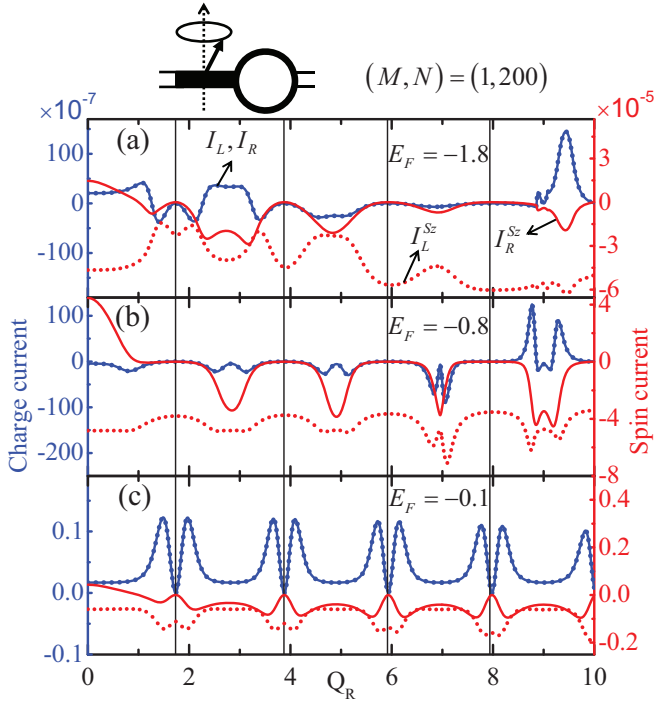,scale=0.44,angle=0,clip=true,trim=0mm 0mm
0mm 0mm}}
\caption{(Color online) Pumped charge and spin currents as a function of the
dimensionless Rashba spin-orbit coupling strength $Q_{R}$ in the
two-terminal spin-driven setup (top schematics) with ring size $%
(M,N)=(1,200) $ at different Fermi energies (a) $E_{F}=-1.8$, (b) $%
E_{F}=-0.8 $, and (c) $E_{F}=-0.1$. The ring is in contact with two
semi-infinite one-dimensional leads in which currents are probed. The solid
vertical lines here indicate the current modulation nodes $Q_{R}^{\ast }$s
at which the complete destructive interferences for spin-$z$ take place,
causing $I_{R}=I_{R}^{S_{z}}=0$. Since charge currents are conserved, $%
I_{L}=I_{R}$ is satisfied. The spin-driven results here correspond to the
electric-driven results, Fig. 3 in Ref.~ \onlinecite{Souma2004}.}
\label{fig2:2LLPM1}
\end{figure}

We begin with the two-terminal case of clean AC ring in the spin-driven
setup Fig. \ref{fig1:setup}(b). Introducing the dimensionless Rashba SOC
strength,
\begin{equation}
Q_{R}\equiv \frac{\gamma _{\text{SO}}N}{\gamma _{0}\pi },  \label{eq:qr}
\end{equation}%
we find in Fig.~\ref{fig2:2LLPM1} for $(M,N)=(1,200)$ and Fig. \ref%
{fig3:2LLPM3_1} for $(M,N)=(3,200)$ where only one channel is open ($M_{%
\text{open}}=1$), the pumped spin-$z$ current $I_{R}^{S_{z}}$ probed by the
right NM lead is a quasi-periodic function of $Q_{R}$; specifically, for $M_{%
\text{open}}=1$, by increasing $Q_{R}$ the $I_{R}^{S_{z}}$ vanishes at
certain $Q_{R}^{\ast }$s, namely, the (AC-spin-interference-induced)
modulation nodes; this behavior is akin to the electric-driven setup where
the charge current disappears at these $Q_{R}^{\ast }$s.\cite%
{Souma2004,Frustaglia2004a,Molnar2004}

\begin{figure}[tbp]
\centerline{\psfig{file=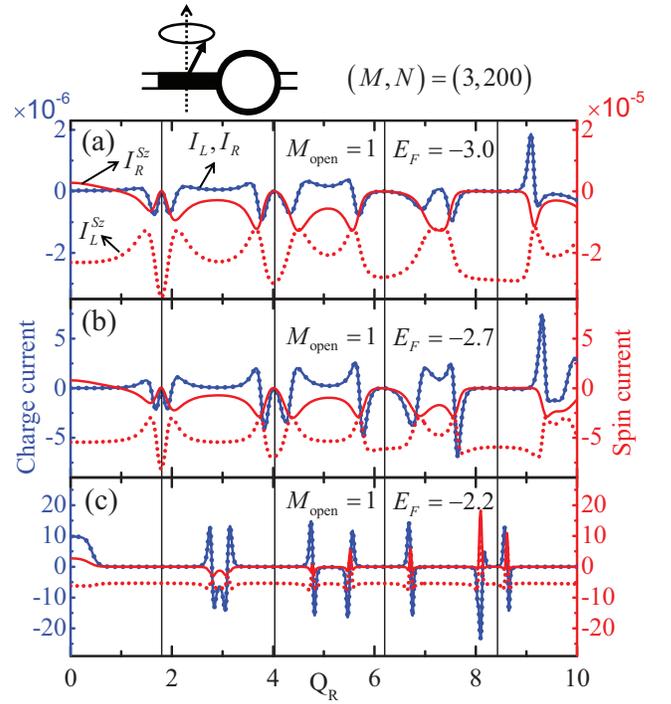,scale=0.43,angle=0,clip=true,trim=0mm 0mm
0mm 0mm}}
\caption{(Color online) Pumped charge and spin currents as a function of the
dimensionless Rashba spin-orbit coupling strength $Q_{R}$ in the
two-terminal spin-driven setup (top schematics) at different Fermi energies
(a) $E_{F}=-3.0$, (b) $E_{F}=-2.7$, and (c) $E_{F}=-2.2$ that yield the
number of open channels, $M_{\text{open}}=1$. The ring is of size $%
(M,N)=(3,200)$ and in contact with two semi-infinite probes of width
consisting of three lattice sites. The solid vertical lines here indicate
the current modulation nodes $Q_{R}^{\ast }$s. The spin-driven results here
correspond to the electric-driven results, Fig. 5 in Ref.~
\onlinecite{Souma2004}.}
\label{fig3:2LLPM3_1}
\end{figure}

The $I_{R}^{S_{z}}$ modulation originates from the fact that a spin-$z$
acquires some AC phase induced by the Rashba SOC when passing through the AC
ring, and the phase difference between the upper-arm and lower-arm of the
ring depends on the Rashba SOC strength; accordingly, gradually varying the
Rashba SOC strength modulates the spin-$z$ current. The condition of $M_{%
\text{open}}=1$ is satisfied when $M=1$ (Fig. \ref{fig2:2LLPM1}), or when $%
M>1$ (Fig. \ref{fig3:2LLPM3_1}) with the Fermi energy $E_{F}$ only crossing
one subband of the leads. In the former (strict 1D), $Q_{R}^{\ast }$s are
independent of the Fermi energy $E_{F}$, while in the latter (quasi 1D),
when one tunes $E_{F}$ but keeps $E_{F}$ in the regime $M_{\text{open}}=1$, $%
Q_{R}^{\ast }$s also remains unaffected (independent of $E_{F}$ as long as $%
M_{\text{open}}=1 $ is satisfied), mimicking again the electric-driven
setup. Moreover, since the spin-$z$ current $I^{S_{z}}$ pumped by the FM is
pure, if the spin-$z$ encounters a complete destructive interference, i.e.,
not able to transport through the ring, then no charge currents will be
generated in the right NM as well, providing that no passage of spins with
different polarizations such as spin-$x$ or spin-$y$ occur through the
interface between the right NM and AC ring as we will address below. We
refer this types of nodes the \emph{AC-spin-interference-induced} modulation
nodes where $I_{R}^{S_{z}}=0$ and $I_{R}$ vanish concurrently at the same $%
Q_{R}^{\ast }$s, as indicated by the solid vertical lines in Figs. \ref%
{fig2:2LLPM1} and \ref{fig3:2LLPM3_1}.

\begin{figure}[tbp]
\centerline{\psfig{file=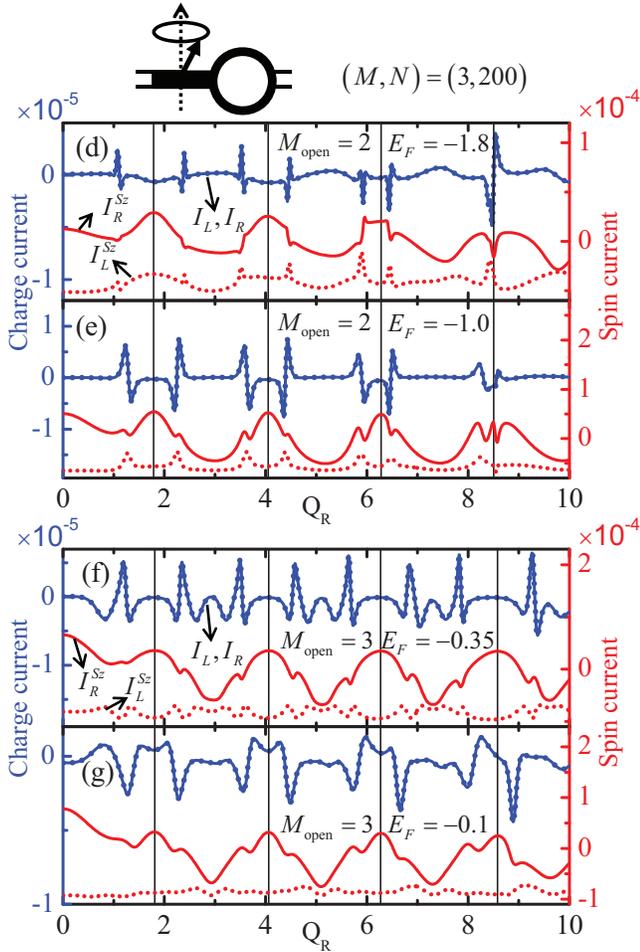,scale=0.45,angle=0,clip=true,trim=0mm 0mm
0mm 0mm}}
\caption{(Color online) Pumped charge and spin currents as a function of the
dimensionless Rashba spin-orbit coupling strength $Q_{R}$ in the spin-driven
setup same as considered in Fig. \protect\ref{fig3:2LLPM3_1}, while
different Fermi energies are chosen to have the number of open channels, $M_{%
\text{open}}=2$ for (d) $E_{F}=-1.8$ and (e) $E_{F}=-1.0$ and $M_{\text{open}%
}=3$ for (f) $E_{F}=-0.35$ and (g) $E_{F}=-0.1$. Unlike (a), (b), and (c)
with $M_{\text{open}}=1$ in Fig. \protect\ref{fig2:2LLPM1}, the modulation
here becomes incomplete (absence of the AC-spin-interference-induced
modulation nodes at which one has $I_{R}=I_{R}^{S_{z}}=0$, namely, absence
of $Q_{R}^{\ast }$s) due to the loss of $M_{\text{open}}=1$, while the
(incomplete) quasi-periodicity can still be found as depicted by the solid
vertical lines. The spin-driven results here correspond to the
electric-driven results, Fig. 6 in Ref. \onlinecite{Souma2004}.}
\label{fig4:2LLPM3_2}
\end{figure}

Note that when a pumped spin-$z$ enters the ring, it starts to precess, and
there are chances for this spin to become spin-$x$ or spin-$y$ when leaving
the ring to the right NM, so that nonzero charge current $I_{R}\neq 0$
without spin-$z$ current $I_{R}^{S_{z}}=0$ can be detected by the right NM.
For $I_{R}^{S_{z}}=0$ nodes involving processes as mentioned above are \emph{%
not} the AC-spin-interference-induced modulation nodes $Q_{R}^{\ast }$s as
being focused here, because they originate from precession but not
interference. Noteworthily, although the 1D ring $(M=1)$ in Fig. \ref%
{fig2:2LLPM1} and quasi-1D ring in Fig. \ref{fig3:2LLPM3_1} are both in the $%
M_{\text{open}}=1$ regime, there is an essential difference between them. In
Fig. \ref{fig2:2LLPM1}, there are no evanescent modes, while in Fig. \ref%
{fig3:2LLPM3_1}, there are two $(M-M_{\text{open}}=3-1)$ channels that
contribute to the evanescent modes; the $Q_{R}^{\ast }$ nodes in Fig. \ref%
{fig3:2LLPM3_1} are thus slightly modified from Fig. \ref{fig2:2LLPM1}.
Also, the $I_{L}^{S_{z}}$ can be nonzero at these $Q_{R}^{\ast }$s, simply
because the FM pumps also spin-$z$ currents directly to the left lead. To
see the corresponding electric-driven results, compare Fig.~\ref{fig2:2LLPM1}
here with Fig. 3 in Ref.~\onlinecite{Souma2004} and Fig.~\ref{fig3:2LLPM3_1}
here with Fig. 5 in Ref.~\onlinecite{Souma2004}.

For $M_{\text{open}}>1$, the conducting spin states become incoherent or
impure due to the entanglements between spin and orbit degrees of freedom.%
\cite{Souma2004} The concept of ensemble average or spin density matrix has
to be introduced to describe the transport of interferences induced by the
AC effect. Furthermore, when more channels are open, the detected spin phase
is obtained by taking into account the transport processes within and
between each single channels, and each transport process gives different AC
phases yielding different interference nodes (places where the complete
destructive interference occurs); thereby, the overall complete destructive
interference is washed out by different transport processes, forming the
\textquotedblleft incomplete\textquotedblright\ modulations (absence of $%
Q_{R}^{\ast }$ nodes); for example, in Fig.~\ref{fig4:2LLPM3_2} for $M_{%
\text{open}}=2$ and $M_{\text{open}}=3$ with $N=200$, although one can still
find the quasi-periodicity as depicted by the solid vertical lines, $I_{R}$
and $I_{R}^{S_{z}}$ in general do not vanish concurrently. In addition, some
of the pumped spins can be reflected in the FM$|$AC-ring interface before
entering the AC ring, so that one has $\left\vert I_{L}^{S_{z}}\right\vert
\gtrsim \left\vert I_{R}^{S_{z}}\right\vert $. To see the electric-driven
case corresponding to Fig. \ref{fig4:2LLPM3_2}, compare Fig. \ref%
{fig4:2LLPM3_2} here with Fig. 6 in Ref. \onlinecite{Souma2004}. Note that
all the above two-terminal results preserve the conservation of charge
currents with $I_{L}=I_{R}$.

\subsection{Single precessing FM island attached to four-terminal mesoscopic
AC ring}

\label{sec:ishe}

\begin{figure}[tbp]
\centerline{\psfig{file=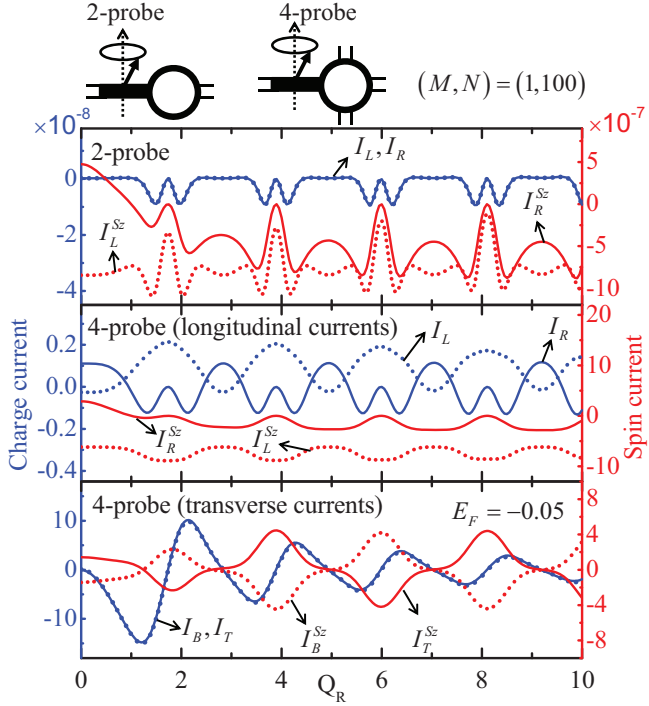,scale=0.445,angle=0,clip=true,trim=0mm 0mm
0mm 0mm}}
\caption{(Color online) Pumped charge and spin currents as a function of the
dimensionless Rashba spin-orbit coupling strength $Q_{R}$ in the
two-terminal (top panel) and four-terminal (middle and bottom panels)
spin-driven setups (see the schematics) in which the same strict
one-dimensional ring of size $(M,N)=(1,100)$ is considered. Each of the
probes is one-dimensional. The complete modulation nodes $Q_{R}^{\ast }$s
that characterize the quasi-periodicity emerge through $%
I_{R}=I_{R}^{S_{z}}=0 $ in both two-and four-terminal setups. In the
four-terminal setup, the transverse currents with $I_{B}=I_{T}$ and $%
I_{B}^{S_{z}}+I_{T}^{S_{z}}=0$ for all $Q_{R}$ reflect the existence of the
inverse spin-Hall effect, and the quasi-periodicity can also be identified
via $I_{B}=I_{T}=0$ at the same $Q_{R}^{\ast }$s. Note that the Onsager
relation is preserved if we compare with the finding in the electric-driven
setup, Fig. 2 in Ref.~ \onlinecite{Souma2005a}.}
\label{fig5:LPM1}
\end{figure}

In the four-terminal setup Fig.~\ref{fig1:setup}(a), both ISHE and AC
effects emerge, giving the inverse quantum-interference-controlled SHE. As
shown in Fig.~\ref{fig5:LPM1} with the ring size $(M,N)=(1,100)$, four
semi-infinite 1D probes, and $E_{F}=-0.05$, the transverse currents obey $%
I_{B}^{S_{z}}=-I_{T}^{S_{z}}$ and $I_{B}=I_{T}$ for all $Q_{R}$, signifying
the ISHE. Figure~\ref{fig5:LPM1} also shows the longitudinal currents in
this four-terminal and the corresponding two-terminal setups. In the
two-terminal setup, again, because of $M_{\text{open}}=1$, the modulation
nodes $Q_{R}^{\ast }$s where $I_{R}$ and $I_{R}^{S_{z}}$ vanish are found,
rendering the complete interference modulation. In the four-terminal setup,
at these $Q_{R}^{\ast }$s, although $I_{R}$ and $I_{R}^{S_{z}}$ vanish as
well, while the longitudinal currents, $I_{L}$ and $I_{L}^{S_{z}}$, do not
vanish, and the inequality $I_{L}\neq I_{R}$ shows up due to the presence of
the top and bottom leads that break the longitudinal current conservation.
Interestingly, the ISHE-induced Hall charge currents $I_{B}=I_{T}$ disappear
at these $Q_{R}^{\ast }$s, which demonstrates again the quasi-periodicity of
the modulation and is dual (satisfies the Onsager relation) to what was
depicted for the SHE-induced Hall spin currents (in the form of spin-Hall
conductance) in Fig. 2 of Ref.~\onlinecite{Souma2005a}. Also note that
transverse currents $I_{B}$ and $I_{T}$ decrease as increasing $Q_{R}$ due
to the reflections in the interfaces between the AC ring and the top or
bottom leads, a manifestation of the lattice Hamiltonian mismatch.

\begin{figure}[tbp]
\centerline{\psfig{file=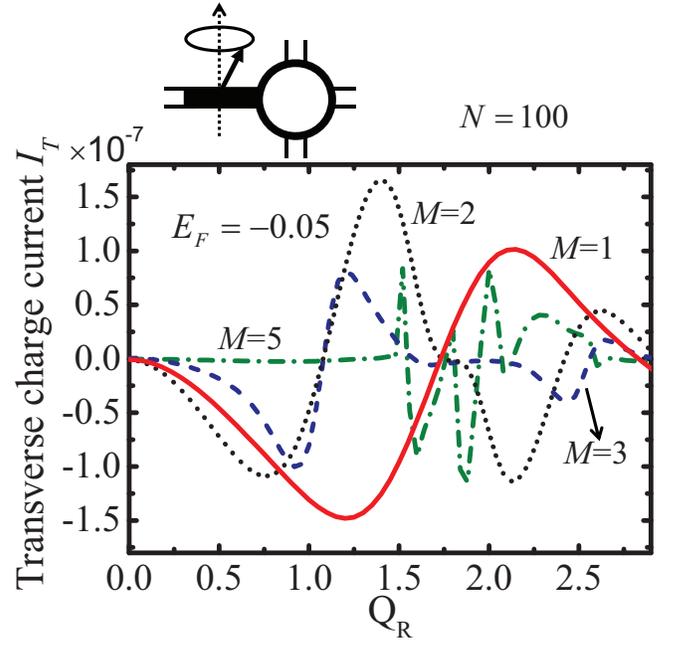,scale=0.44,angle=270,clip=true,trim=0mm 0mm
0mm 0mm}}
\caption{(Color online) Pumped transverse charge currents $I_{T}=I_{B}$ at
different ring width $M$ as a function of the dimensionless Rashba
spin-orbit coupling strength $Q_{R}$ in the four-terminal spin-driven setup
(top schematics) with four semi-infinite one-dimensional probes. Refer to
Fig. 3 in Ref. \onlinecite{Souma2005a} for the reciprocal Onsager
(electric-driven) results.}
\label{fig6:4LLPM}
\end{figure}

To see how the width of the ring $M$ affects the interference, we consider
the AC ring with fixed $N=100$ in contact with four semi-infinite 1D leads
such that only one channel is available for transport in each lead, i.e., $%
M_{\text{open}}=1$. Figure \ref{fig6:4LLPM} indicates that the modulation
frequency of the Hall currents $I_{B}=I_{T}$ for $M=2$ is almost double to
that for $M=1$, because in $M=2$, one additional transport ring path
appears. For larger width, the oscillations of Hall currents become vague
since the multiple intertwined 1D ring paths smear out the periodic behavior
of the currents or average over the AC phase; nevertheless, the complete
quasi-periodicity ($I_{B}=I_{T}=0$ at current modulation nodes $Q_{R}^{\ast
} $s) is protected by the $M_{\text{open}}=1$ condition. The reciprocal
features (for the corresponding electric-driven setup) are shown in Fig. 3
of Ref. \onlinecite{Souma2005a}). Note that the ISHE emerging through $%
I_{B}^{S_{z}}=-I_{T}^{S_{z}}$ and $I_{B}=I_{T}$ is still preserved robustly
against the ring width.

\begin{figure}[tbp]
\centerline{\psfig{file=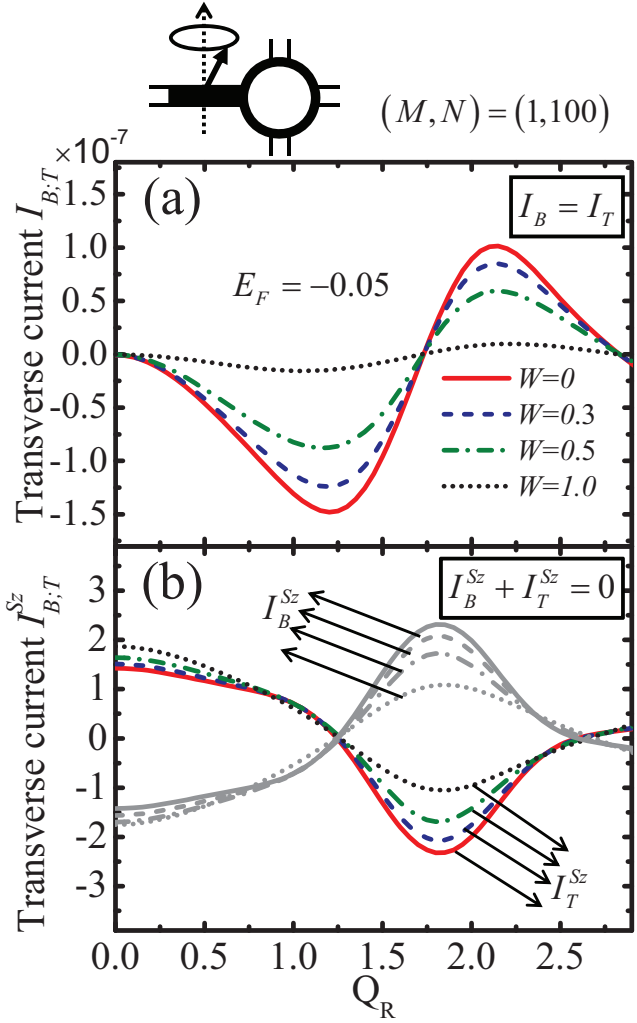,scale=0.48,angle=0,clip=true,trim=0mm 0mm
0mm 0mm}}
\caption{(Color online) Pumped transverse (a) charge and (b) spin-$z$
currents versus the dimensionless Rashba spin-orbit coupling strength $Q_{R}$%
. The inverse spin-Hall effect (shown through $I_{B}=I_{T}$ and $%
I_{B}^{S_{z}}+I_{T}^{S_{z}}=0$) survives in the weak disorder regime with
different disorder strengths $W$ of the ring in the four-terminal (each of
the probes is one-dimensional) spin-driven setup (top schematics). The
reciprocal Onsager (electric-driven) result is shown in Fig. 4 of Ref.
\onlinecite{Souma2005a}.}
\label{fig7:dis_4LLPM}
\end{figure}

Interestingly, the ISHE remains unaffected even in the weak disorder regime.
Figure \ref{fig7:dis_4LLPM} plots the probed currents with different (weak)
disorder strength $W$ modeled by the random on-site potentials of the ring,
namely, $\varepsilon ^{n,m}\in \lbrack -W/2,W/2]$. The modulations of $%
I_{B}^{S_{z}}=-I_{T}^{S_{z}}$ and $I_{B}=I_{T}$ show that ISHE is robust
against week disorder. In addition, the presence of the week disorder plays
merely the role to reduce the amplitudes of the modulation as also addressed
in Ref. \onlinecite{Souma2005a} for the corresponding electric-driven setup
(compare Fig. \ref{fig7:dis_4LLPM} here with Fig. 4 in Ref. %
\onlinecite{Souma2005a}).

\subsection{Symmetry operations relating pumped spin and charge currents}

\label{sec:sysinv}

\begin{figure}[tbp]
\centerline{\psfig{file=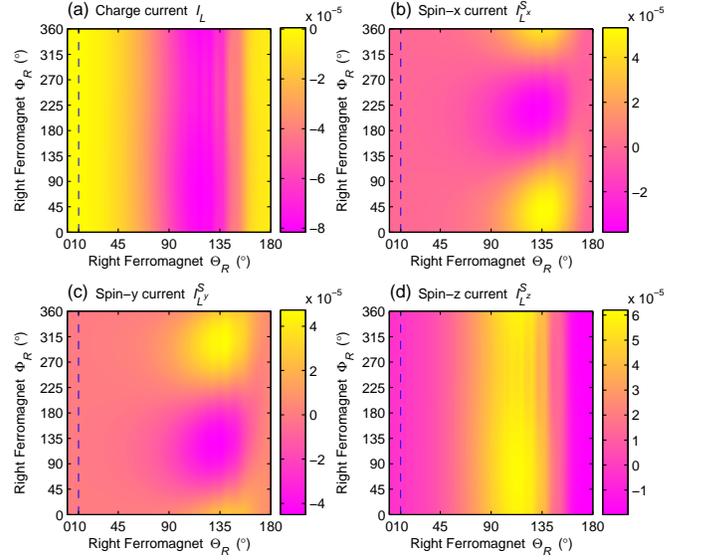,scale=0.52,angle=0,clip=true,trim=0mm 0mm
0mm 0mm}}
\caption{(Color online) Pumped (a) charge current and (b) spin-$x$, (c) spin-%
$y$, and (d) spin-$z$ currents, in the two-terminal
two-precessing-ferromagnet ring device as shown in the schematics of Fig.~%
\protect\ref{fig10:LPRP}, probed by the left lead as functions of precession
cone angle $\Theta _{R}$ and initial precession phase (azimuthal angle) $%
\Phi _{R}$ of the right ferromagnet. The ring is of size $(M,N)=(1,200)$ and
of the dimensionless Rashba spin-orbit coupling strength $Q_{R}=5$. The left
ferromagnet is of $\Theta _{L}=10^{\circ }$ (represented by the dash lines)
and $\Phi _{L}=0^{\circ }$. The left and right leads are one-dimensional,
i.e., the number of open channel $M_{\text{open}}=1$ with Fermi-energy $%
E_{F}=-1.8$.}
\label{fig8:2FM_Lcu}
\end{figure}

We extend our study to the case of multiple precessing FM islands and
examine the relations between the pumped currents. Consider the two-terminal
setup Fig. \ref{fig1:setup}(b) with an additional FM island inserted between
the ring and the right NM (as the schematics shown in Fig.~\ref{fig10:LPRP}%
). Let $\Theta _{R}$ ($\Theta _{L}$) be the precession cone angle and $\Phi
_{R}$ ($\Phi _{L}$) be the initial precession phase of the right (left) FM.
For $M_{\text{open}}=1$ at the condition $Q_{R}\approx Q_{R}^{\ast }$, we
find that the spin-$z$ currents probed by the left (right) lead remain
almost constant when varying $\Theta _{R}$ ($\Theta _{L}$) and/or $\Phi _{R}$
($\Phi _{L}$); in other words, the left FM does not communicate with the
right FM due to the complete destructive interference. Contrarily, in Fig. %
\ref{fig8:2FM_Lcu}, with $E_{F}=-1.8$, $(M,N)=(1,200)$ ring, two (left and
right) 1D leads, fixed $\Theta _{L}=10^{\circ }$ (indicated by the dash
line) and $\Phi _{L}=0^{\circ }$ in the left FM, at $Q_{R}=5$, i.e., the
condition of the complete destructive interference is off, we see that the
pumped currents probed by the left lead, $I_{L}$, $I_{L}^{S_{x}}$, $%
I_{L}^{S_{y}}$, and $I_{L}^{S_{z}}$, significantly changes with $\Theta _{R}$%
. For $\Phi _{R}$ dependence, noteworthily, we see that $I_{L}$ and $%
I_{L}^{S_{z}}$ are not as sensitive to $\Phi _{R}$ as $I_{L}^{S_{x}}$ and $%
I_{L}^{S_{y}}$ (for example, compare subfigures of Fig. \ref{fig8:2FM_Lcu}
at $\Theta _{R}\approx 135^{\circ }$).

\begin{figure}[tbp]
\centerline{\psfig{file=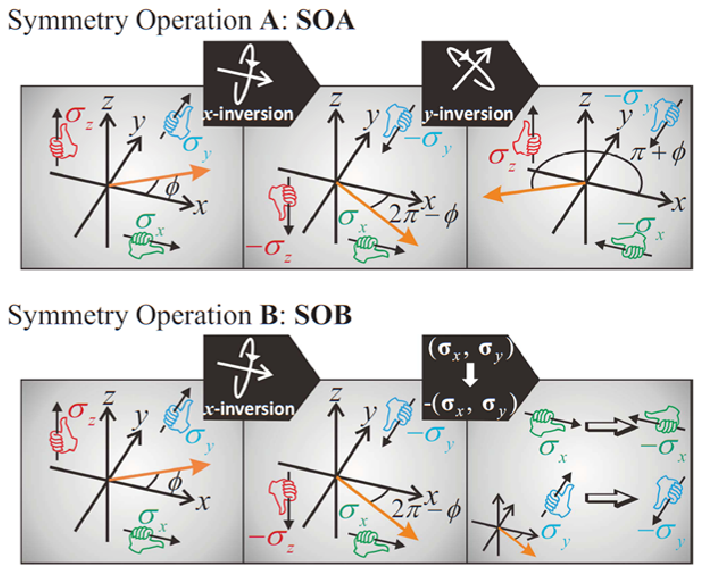,scale=1.2,angle=270,clip=true,trim=0mm 0mm
135mm 182mm}}
\caption{(Color online) Symmetry operations \textbf{A} (SOA) and \textbf{B }%
(SOB). In SOA, the inversion with respect to the $+x$ axis (as depicted by
the inset, $x$-inversion) is first performed to the system represented by
the orange/gray arrow lying on the $x$-$y$ plane and then with respect to $%
+y $ axis (as depicted by the inset, $y$-inversion). In SOB, $x$-inversion
is first performed and then the replacement $(\protect\sigma _{x},\protect%
\sigma _{y})\rightarrow -(\protect\sigma _{x},\protect\sigma _{y})$. The
successive figures show how the spin $(\protect\sigma _{x},\protect\sigma %
_{y},\protect\sigma _{z})$ and polar angle $\protect\phi $ change after each
inverting or replacing.}
\label{fig9:SOs}
\end{figure}

To establish a systematic analysis on the relations between pumped currents,
we inspect the device from the symmetry perspective. Recall that the Rashba
SOC originates from the structural inversion asymmetry,\cite{Rashba1960}
meaning that the AC ring Hamiltonian, Eq. (\ref{eq:H_RR}) defined by Eq. (%
\ref{eq:gamma_phi}) and Eq. (\ref{eq:gamma_r}), does not remain the same by
inverting the ring once. This one-time inversion asymmetry, however, leads
to the conjecture that if one can somehow perform some inversion-like
operations twice, then $H_{\text{ACR}}$ might be invariant. Indeed, at least
two types of symmetry operations can render invariant $H_{\text{ACR}}$.
These two operations are illustrated in Fig. \ref{fig9:SOs}. We refer the
first operation as symmetry operation \textbf{A} (abbreviated as SOA), and
the second as symmetry operation \textbf{B }(abbreviated as SOB). In SOA, we
first invert the system with respect to the $+x$ axis ($x$-inversion) and
then invert again with respect to the $+y$ axis ($y$-inversion); note that
each inversion gives a $\Delta \phi \rightarrow -\Delta \phi $ and a $\sigma
_{z}\rightarrow -\sigma _{z}$. After these two inversions, as shown in Fig. %
\ref{fig9:SOs} we have, $\phi \rightarrow \pi +\phi $, $\Delta \phi
\rightarrow \Delta \phi $, and
\begin{equation}
(\sigma _{x},\sigma _{y},\sigma _{z})\rightarrow (-\sigma _{x},-\sigma
_{y},\sigma _{z})\text{, for SOA}  \label{eq:SOA_sigma}
\end{equation}%
such that Eqs. (\ref{eq:gamma_phi}) and (\ref{eq:gamma_r}) remain unaltered,
conceding invariant $H_{\text{ACR}}$. For SOB, we first perform the $x$%
-inversion and then the replacement $(\sigma _{x},\sigma _{y})\rightarrow $ $%
-(\sigma _{x},\sigma _{y})$; note that in SOB, the $\sigma _{y}$ undergoes $%
\sigma _{y}\rightarrow -\sigma _{y}\rightarrow \sigma _{y}$ ($\sigma
_{y}\rightarrow -\sigma _{y}$ due to the inversion and then $-\sigma
_{y}\rightarrow \sigma _{y}$ due to the replacement). We thus get (refer to
Fig. \ref{fig9:SOs}), $\phi \rightarrow 2\pi -\phi $, $\Delta \phi
\rightarrow -\Delta \phi $, and
\begin{equation}
(\sigma _{x},\sigma _{y},\sigma _{z})\rightarrow (-\sigma _{x},\sigma
_{y},-\sigma _{z})\text{, for SOB}  \label{eq:SOB_sigma}
\end{equation}%
so that Eqs. (\ref{eq:gamma_phi}) and (\ref{eq:gamma_r}) are unchanged to
yield invariant $H_{\text{ACR}}$ as well.

For NMs, obviously, after SOA or SOB, the $H_{\text{NM}}$ remains the same,
because all NMs are of the same spin-independent
structural-inversion-invariant Hamiltonian. Also, all the hybridizations
(characterized by the same spin-independent hopping $-\gamma _{0}$), $H_{%
\text{NM-ACR}}$, $H_{\text{FM-ACR}}$, and $H_{\text{FM-NM}}$ are SOA- and
SOB-invariant. The only portion in the total Hamiltonian that might not be
able to recover to its original form is the time-dependent Hamiltonian $H_{%
\text{FM}}\left( t\right) $, because under SOA or SOB the directions of the
precession axis can vary. However, note that since what we are interested in
is the \emph{time-averaged} currents, it is the \emph{relative} initial
precessing phase $\Phi _{L}-\Phi _{R}$ that is relevant to these average
currents, while the $\Phi _{L}-\Phi _{R}$ does not change under SOA nor SOB,
because SOA or SOB are applied to the whole system (i.e., to all precessing
FMs). Moreover, any operations or transformations will transfer one pumping
configuration either to the same configuration or to another configuration;
in the former, the symmetry argument will relate the probed pumped currents
within a single configuration, whereas in the later, relate the probed
currents between two different configurations; this will become more clear
in the next section. Without loss of generality, in what follows, we choose
systems originally at $\Phi _{L}=\Phi _{R}=0^{\circ }$ to illustrate how SOA
or SOB helps construct the relations between different probed currents and
verify these relations by inspecting our numerical results.

\subsection{Symmetry arguments applied to two precessing FM islands with
two-terminal mesoscopic AC ring in between}

\label{sec:sys_nresults}
\begin{figure}[tbp]
\centerline{\psfig{file=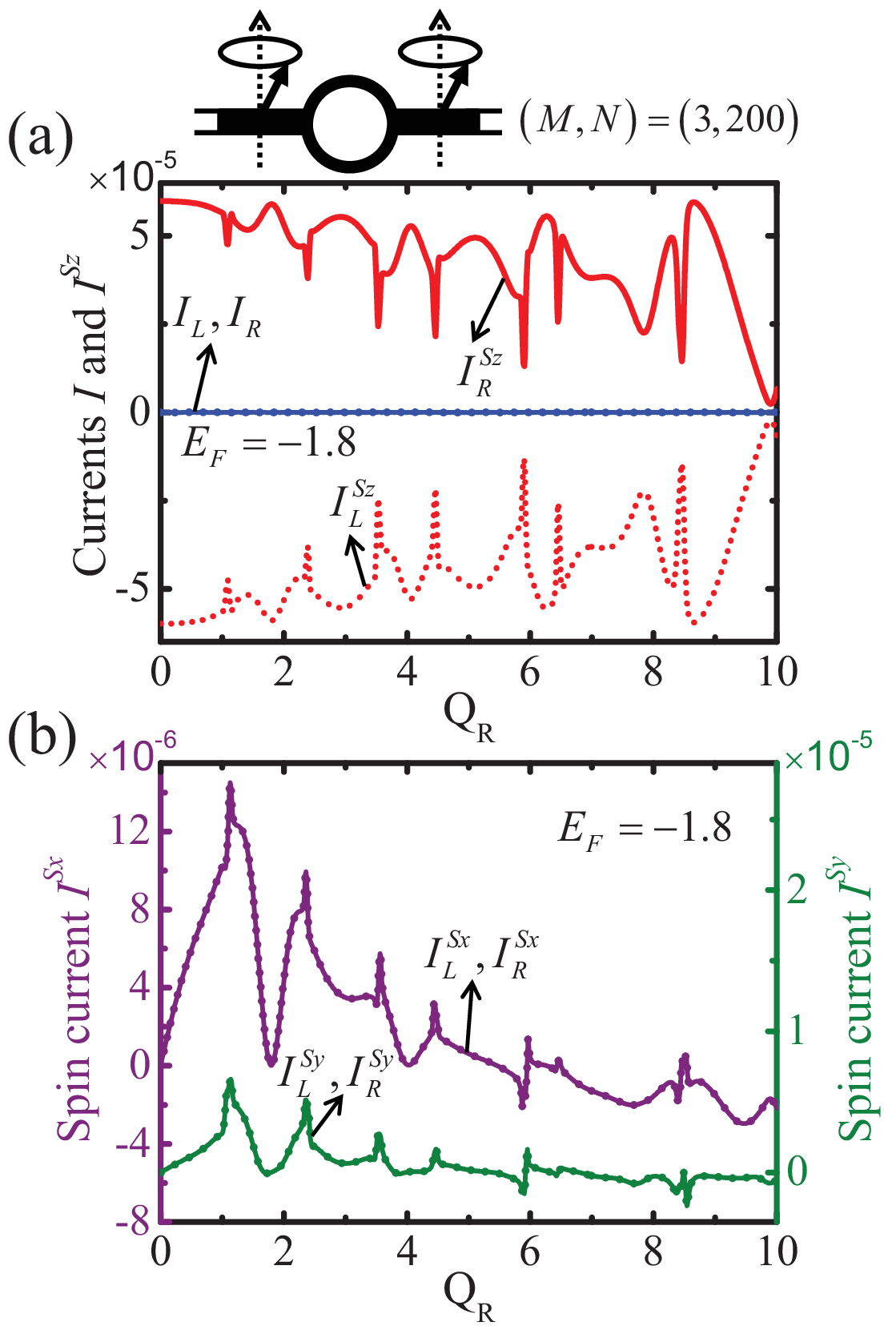,scale=0.45,angle=0,clip=true,trim=0mm 0mm
0mm 0mm}}
\caption{(Color online) Pumped (a) charge and spin-$z$ currents and (b) spin-%
$x$ and spin-$y$ currents probed by the left and right leads of finite width
consisting of three lattice points versus the dimensionless Rashba
spin-orbit coupling strength $Q_{R}$ in the two-terminal two-precessing-FM
setup $P_{10^{\circ }}$-$P_{10^{\circ }}$ (top schematics), i.e., the left
ferromagnet and right ferromagnet are of precession cone angle $10^{\circ }$
and precession axes both along $+z$ direction.}
\label{fig10:LPRP}
\end{figure}

For the purpose of demonstration, we choose to consider here the
two-terminal two-precessing-FM (left and right FMs adjacent to the left and
to the right of the ring, respectively) setups, while one can apply the
argument presented below also to the ring devices consist of arbitrary
number of terminals and precessing FM islands. In addition, since the
definitions of SOA and SOB have nothing to do with the ring width $M$, ring
length $N$, number of open channels $M_{\text{open}}$, and $E_{F}$, \emph{%
the symmetry argument is valid for any }$M$\emph{, }$N$\emph{, }$M_{\text{%
open}}$\emph{, and }$E_{F}$. Here, we choose $\left( M,N\right) =(3,200)$
and set $E_{F}=-1.8\gamma _{0}$ giving $M_{\text{open}}=2$ to exemplify the
symmetry operations. We use the notation convention $A$-$B$ to describe the
pumping configuration, with $A$ accounting for the left precessing FM and $B$
for the right precessing FM. Here, with $\{A$, $B\}\in \{P_{\Theta }$, $\bar{%
P}_{\Theta }\}$, $P_{\Theta }$ ($\bar{P}_{\Theta }$) stands for the FM that
is of precession axis along $+z$ ($-z$) axis and of precession cone angle $%
\Theta $. For example, the schematics in Fig. \ref{fig10:LPRP} is notated as
$P_{10^{\circ }}$-$P_{10^{\circ }}$, in Fig. \ref{fig12:LPRAP} as $%
P_{10^{\circ }}$-$\bar{P}_{10^{\circ }}$, and in Fig. \ref{fig17:H_LAPRP} as
$\bar{P}_{90^{\circ }}$-$P_{90^{\circ }}$.

\begin{figure}[tbp]
\centerline{\psfig{file=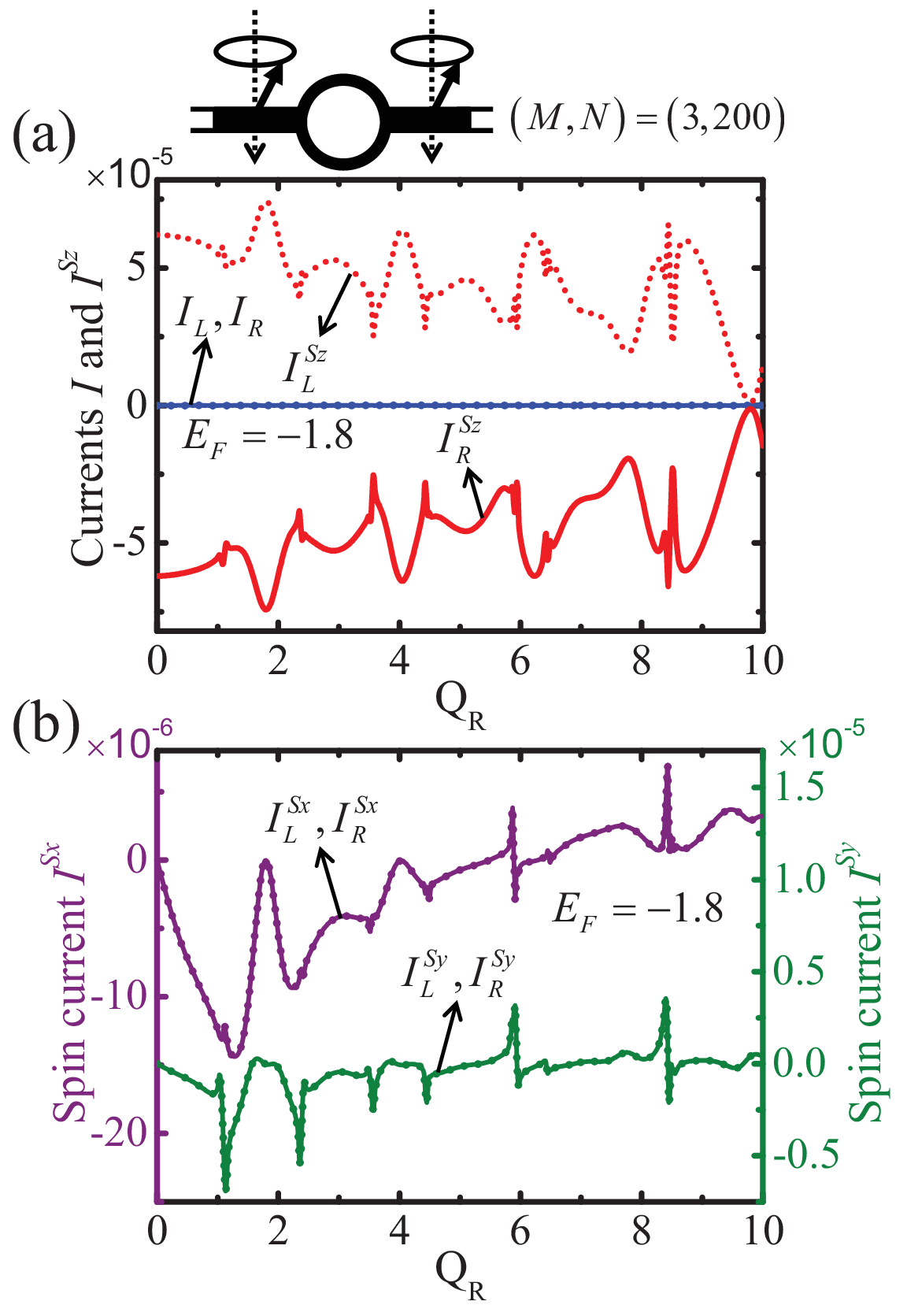,scale=0.45,angle=0,clip=true,trim=0mm 0mm
0mm 0mm}}
\caption{(Color online) Pumped (a) charge and spin-$z$ currents and (b) spin-%
$x$ and spin-$y$ currents versus the dimensionless Rashba spin-orbit
coupling strength $Q_{R}$ in the two-precessing-FM setup $\bar{P}_{10^{\circ
}}$-$\bar{P}_{10^{\circ }}$ (top schematics) same as the one considered in
Fig. \protect\ref{fig10:LPRP} but with precession axes along $-z$ direction
for both the left and right ferromagnets.}
\label{fig11:LAPRAP}
\end{figure}

\begin{figure}[tbp]
\centerline{\psfig{file=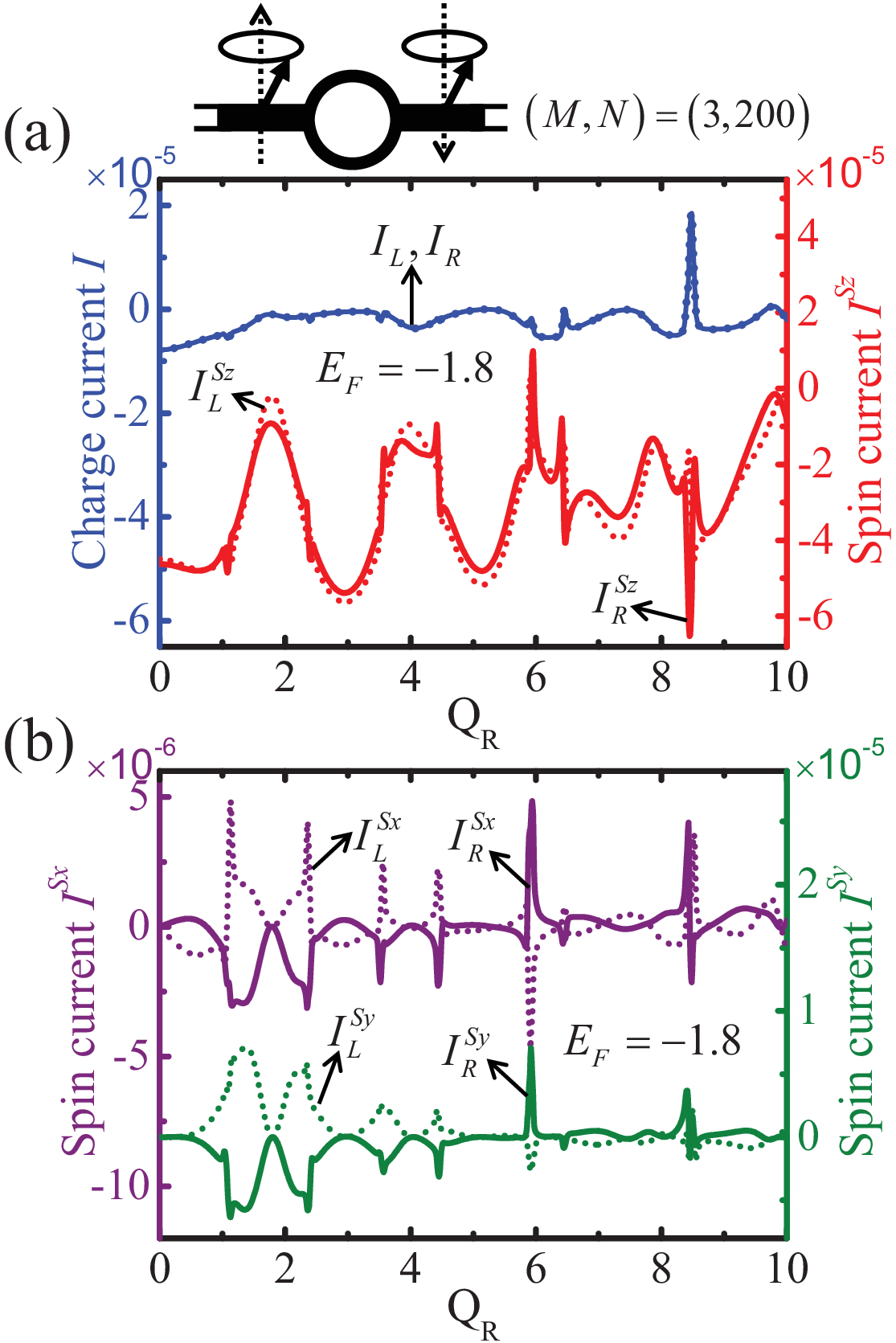,scale=0.45,angle=0,clip=true,trim=0mm 0mm
0mm 0mm}}
\caption{(Color online) Pumped (a) charge and spin-$z$ currents and (b) spin-%
$x$ and spin-$y$ currents versus the dimensionless Rashba spin-orbit
coupling strength $Q_{R}$ in the two-precessing-FM setup $P_{10^{\circ }}$-$%
\bar{P}_{10^{\circ }}$ (top schematics) same as the one considered in Fig.
\protect\ref{fig10:LPRP} but with precession axis along $+z$ ($-z$)
direction for the left (right) ferromagnet.}
\label{fig12:LPRAP}
\end{figure}

\begin{figure}[tbp]
\centerline{\psfig{file=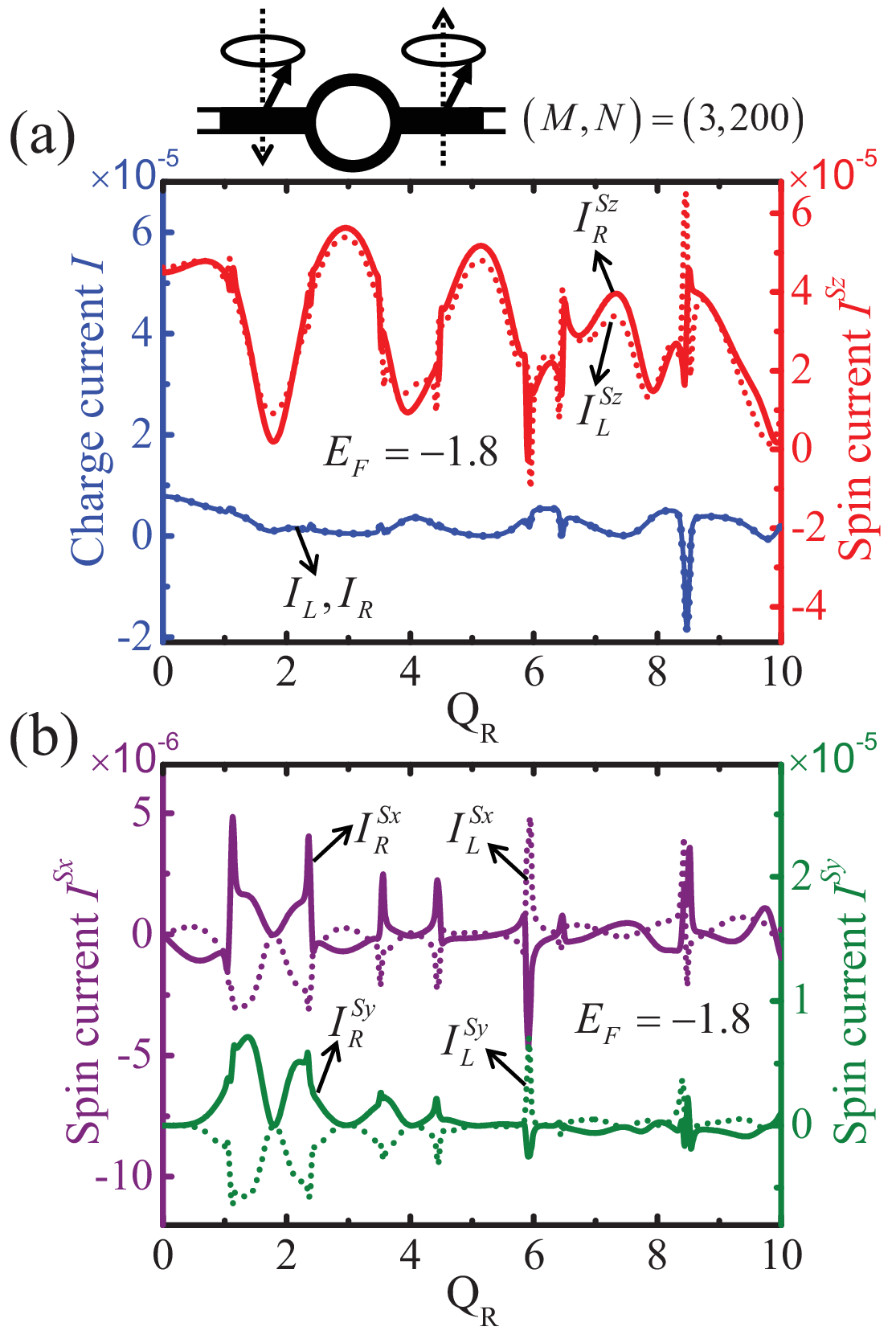,scale=0.45,angle=0,clip=true,trim=0mm 0mm
0mm 0mm}}
\caption{(Color online) Pumped (a) charge and spin-$z$ currents and (b) spin-%
$x$ and spin-$y$ currents versus the dimensionless Rashba spin-orbit
coupling strength $Q_{R}$ in the two-precessing-FM setup $\bar{P}_{10^{\circ
}}$-$P_{10^{\circ }}$ (top schematics) same as the one considered in Fig.
\protect\ref{fig10:LPRP} but with precession axis along $-z$ ($+z$)
direction for the left (right) ferromagnet.}
\label{fig13:LAPRP}
\end{figure}

Focus on SOA first. Consider $P_{10^{\circ }}$-$P_{10^{\circ }}$, Fig. \ref%
{fig10:LPRP}. By applying SOA to $P_{10^{\circ }}$-$P_{10^{\circ }}$, the
first step ($x$-inversion) generates $\bar{P}_{80^{\circ }}$-$\bar{P}%
_{80^{\circ }}$, while the second step ($y$-inversion) yields the swap $L$
(left) $\leftrightarrow R$ (right) and turns $\bar{P}_{80^{\circ }}$-$\bar{P}%
_{80^{\circ }}$ into $P_{10^{\circ }}$-$P_{10^{\circ }}$, i.e., the original
pumping configuration is recovered. As a result, in $P_{10^{\circ }}$-$%
P_{10^{\circ }}$ we have, due to the $y$-inversion involved in SOA, $%
I_{L}=-I_{R}$ (or $I_{L}^{S_{q}}=-I_{R}^{S_{q}}$ for all $q$ components
before any operations on spins), which then incorporated with the
replacement (\ref{eq:SOA_sigma}) turns $I_{L}^{S_{q}}=-I_{R}^{S_{q}}$ into $%
(I_{L}^{S_{x}},I_{L}^{S_{y}},I_{L}^{S_{z}})=(I_{R}^{S_{x}},I_{R}^{S_{y}},-I_{R}^{S_{z}})
$, in line with our numerical result, Fig. \ref{fig10:LPRP}. In Fig. \ref%
{fig11:LAPRAP} ($\bar{P}_{10^{\circ }}$-$\bar{P}_{10^{\circ }}$), by
employing the same argument based on SOA, we obtain again the relations $%
I_{L}=-I_{R}$ and $%
(I_{L}^{S_{x}},I_{L}^{S_{y}},I_{L}^{S_{z}})=(I_{R}^{S_{x}},I_{R}^{S_{y}},-I_{R}^{S_{z}})
$. Being noteworthy, in the pumping configuration $A$-$A$ such as Figs. \ref%
{fig10:LPRP} and \ref{fig11:LAPRAP}, the probed charge currents vanish $%
I_{L}=I_{R}=0$ due to the left-right transmission symmetry,\cite{Chen2009}
resulting in\emph{\ pure }spin currents in the NMs; this absence of charge
currents can also be obtained by noting that the current conservation $%
I_{L}=I_{R}$ and the symmetry argument that gives $I_{L}=-I_{R}$ have to be
satisfied simultaneously.

\begin{figure}[tbp]
\centerline{\psfig{file=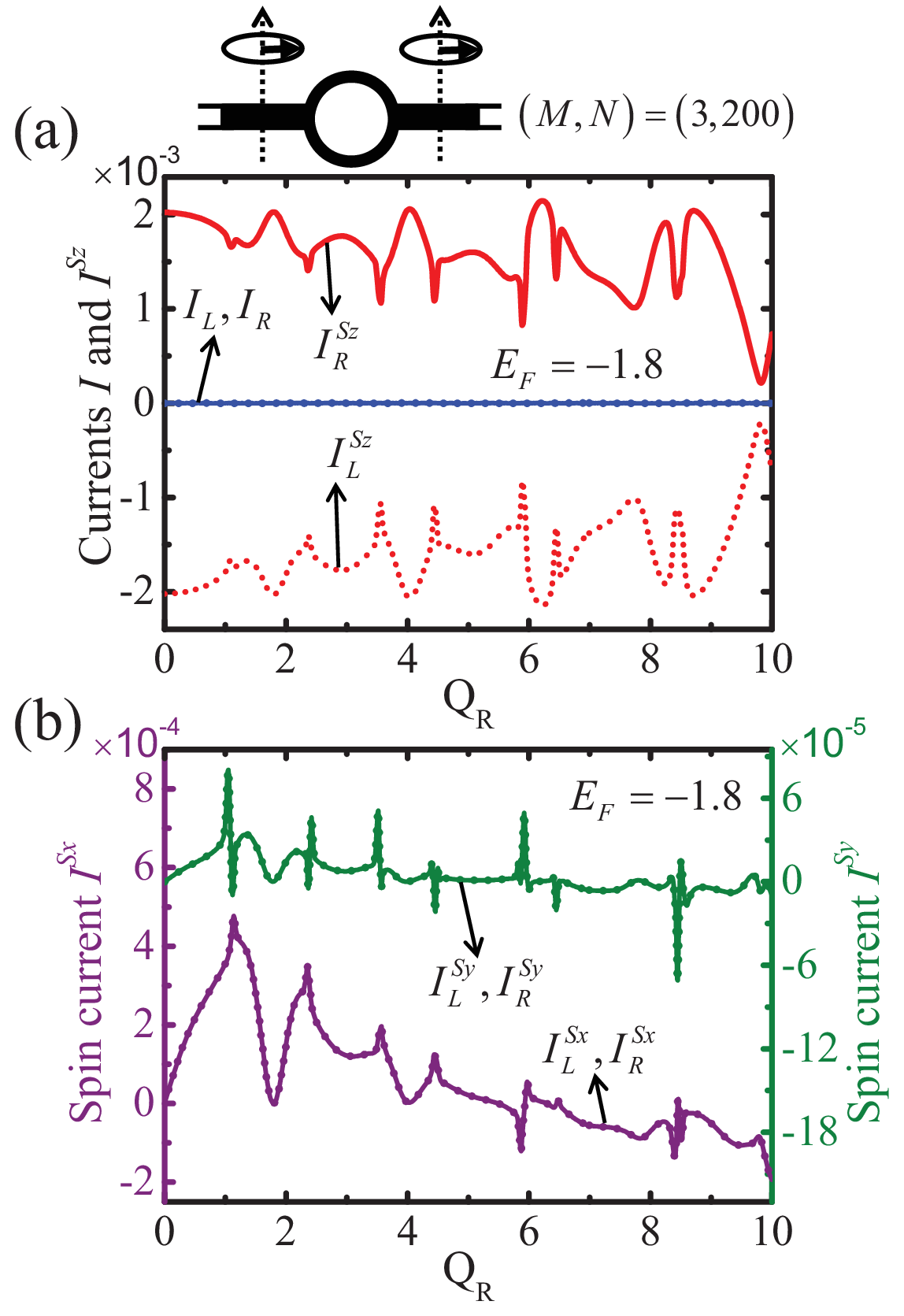,scale=0.45,angle=0,clip=true,trim=0mm 0mm
0mm 0mm}}
\caption{(Color online) Pumped (a) charge and spin-$z$ currents and (b) spin-%
$x$ and spin-$y$ currents probed by the left and right leads of finite width
consisting of three lattice points versus the dimensionless Rashba
spin-orbit coupling strength $Q_{R}$ in the two-terminal two-precessing-FM
setup $P_{90^{\circ }}$-$P_{90^{\circ }}$ (top schematics), i.e., both the
left and right ferromagnets are of precession cone angle $90^{\circ }$ and
precession axis along $+z$ direction.}
\label{fig14:H_LPRP}
\end{figure}

On the other hand, for the left-right transmission asymmetric cases such as
Figs. \ref{fig12:LPRAP} ($P_{10^{\circ }}$-$\bar{P}_{10^{\circ }}$) and \ref%
{fig13:LAPRP} ($\bar{P}_{10^{\circ }}$-$P_{10^{\circ }}$), $I_{L}$ and $%
I_{R} $ in general can be nonzero. Similarly, performing SOA on $%
P_{10^{\circ }}$-$\bar{P}_{10^{\circ }}$ ($\bar{P}_{10^{\circ }}$-$%
P_{10^{\circ }}$), the $x$-inversion renders $\bar{P}_{80^{\circ }}$-$%
P_{80^{\circ }}$ ($P_{80^{\circ }}$-$\bar{P}_{80^{\circ }}$), and then the
proceeding $y$-inversion gives $\bar{P}_{10^{\circ }}$-$P_{10^{\circ }}$ ($%
P_{10^{\circ }}$-$\bar{P}_{10^{\circ }}$); hence, SOA transfers $%
P_{10^{\circ }}$-$\bar{P}_{10^{\circ }}$ ($\bar{P}_{10^{\circ }}$-$%
P_{10^{\circ }}$) to the different pumping configuration $\bar{P}_{10^{\circ
}}$-$P_{10^{\circ }}$ ($P_{10^{\circ }}$-$\bar{P}_{10^{\circ }}$).
Therefore, the current $I_{L;R}$ in $P_{10^{\circ }} $-$\bar{P}_{10^{\circ
}} $ ($\bar{P}_{10^{\circ }}$-$P_{10^{\circ }}$) equals to $-I_{R;L}$ in $%
\bar{P}_{10^{\circ }}$-$P_{10^{\circ }}$ ($P_{10^{\circ }}$-$\bar{P}%
_{10^{\circ }}$). Again, the relations above for charge currents together
with the replacement (\ref{eq:SOA_sigma}) make $%
(I_{L}^{S_{x}},I_{L}^{S_{y}},I_{L}^{S_{z}})$ in $P_{10^{\circ }}$-$\bar{P}%
_{10^{\circ }}$ ($\bar{P}_{10^{\circ }}$-$P_{10^{\circ }}$) identical to $%
(I_{R}^{S_{x}},I_{R}^{S_{y}},-I_{R}^{S_{z}})$ in $\bar{P}_{10^{\circ }}$-$%
P_{10^{\circ }}$ ($P_{10^{\circ }}$-$\bar{P}_{10^{\circ }}$) and $%
(I_{R}^{S_{x}},I_{R}^{S_{y}},I_{R}^{S_{z}})$ in $P_{10^{\circ }}$-$\bar{P}%
_{10^{\circ }}$ ($\bar{P}_{10^{\circ }}$-$P_{10^{\circ }}$) identical to $%
(I_{L}^{S_{x}},I_{L}^{S_{y}},-I_{L}^{S_{z}})$ in $\bar{P}_{10^{\circ }}$-$%
P_{10^{\circ }}$ ($P_{10^{\circ }}$-$\bar{P}_{10^{\circ }}$). All above
features are again in line with our numerical results, Fig. \ref{fig12:LPRAP}
and Fig. \ref{fig13:LAPRP}. It should be noted here that $\Theta =10^{\circ
} $ in our numerical calculation is chosen merely for the illustrations of
symmetry operations. The symmetry argument presented above in fact is
applicable for any cone angle $\Theta $ and even for the case of $\Theta
_{L}\neq \Theta _{R}$, i.e., the precessing cone angles for left ($\Theta
_{L}$) and right ($\Theta _{R}$) FMs are different. Particularly, at $\Theta
=\Theta _{L}=\Theta _{R}=90^{\circ }$, all the relations based on SOA shown
above are preserved as well (see Figs. \ref{fig14:H_LPRP}, \ref%
{fig15:H_LAPRAP}, \ref{fig16:H_LPRAP}, and \ref{fig17:H_LAPRP}), while since
the pumping configuration for $\Theta =90^{\circ }$ (precession within the
two-dimensional $x$-$y$ plane) is of higher geometrical symmetry than $%
\Theta =10^{\circ }$, the probed currents are of additional relations as
demonstrated below.

\begin{figure}[tbp]
\centerline{\psfig{file=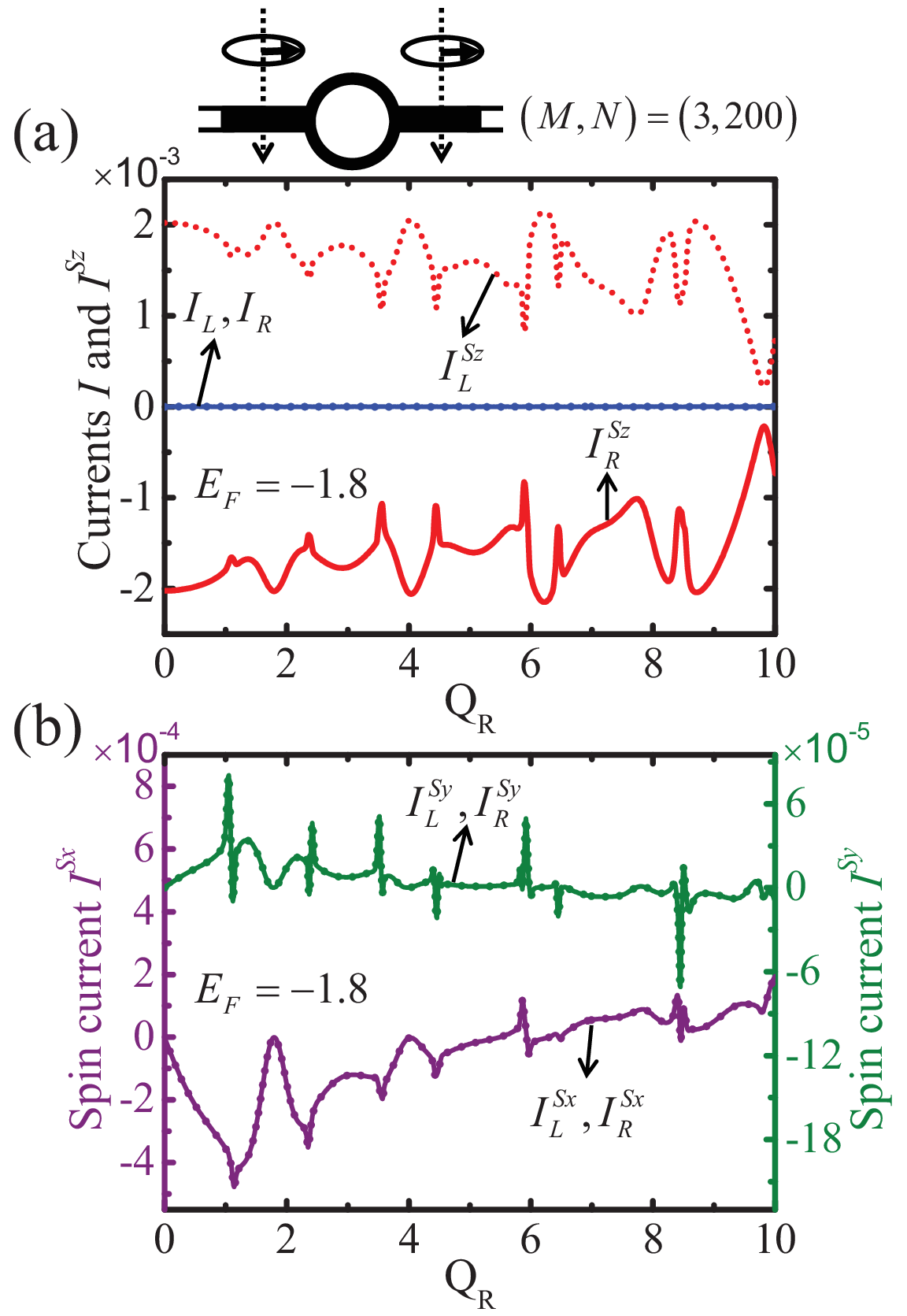,scale=0.45,angle=0,clip=true,trim=0mm 0mm
0mm 0mm}}
\caption{(Color online) Pumped (a) charge and spin-$z$ currents and (b) spin-%
$x$ and spin-$y$ currents versus the dimensionless Rashba spin-orbit
coupling strength $Q_{R}$ in the two-precessing-FM setup $\bar{P}_{90^{\circ
}}$-$\bar{P}_{90^{\circ }}$ (top schematics) same as the one considered in
Fig. \protect\ref{fig14:H_LPRP} but with precession axis along $-z$
direction for both left and right ferromagnets.}
\label{fig15:H_LAPRAP}
\end{figure}

\begin{figure}[tbp]
\centerline{\psfig{file=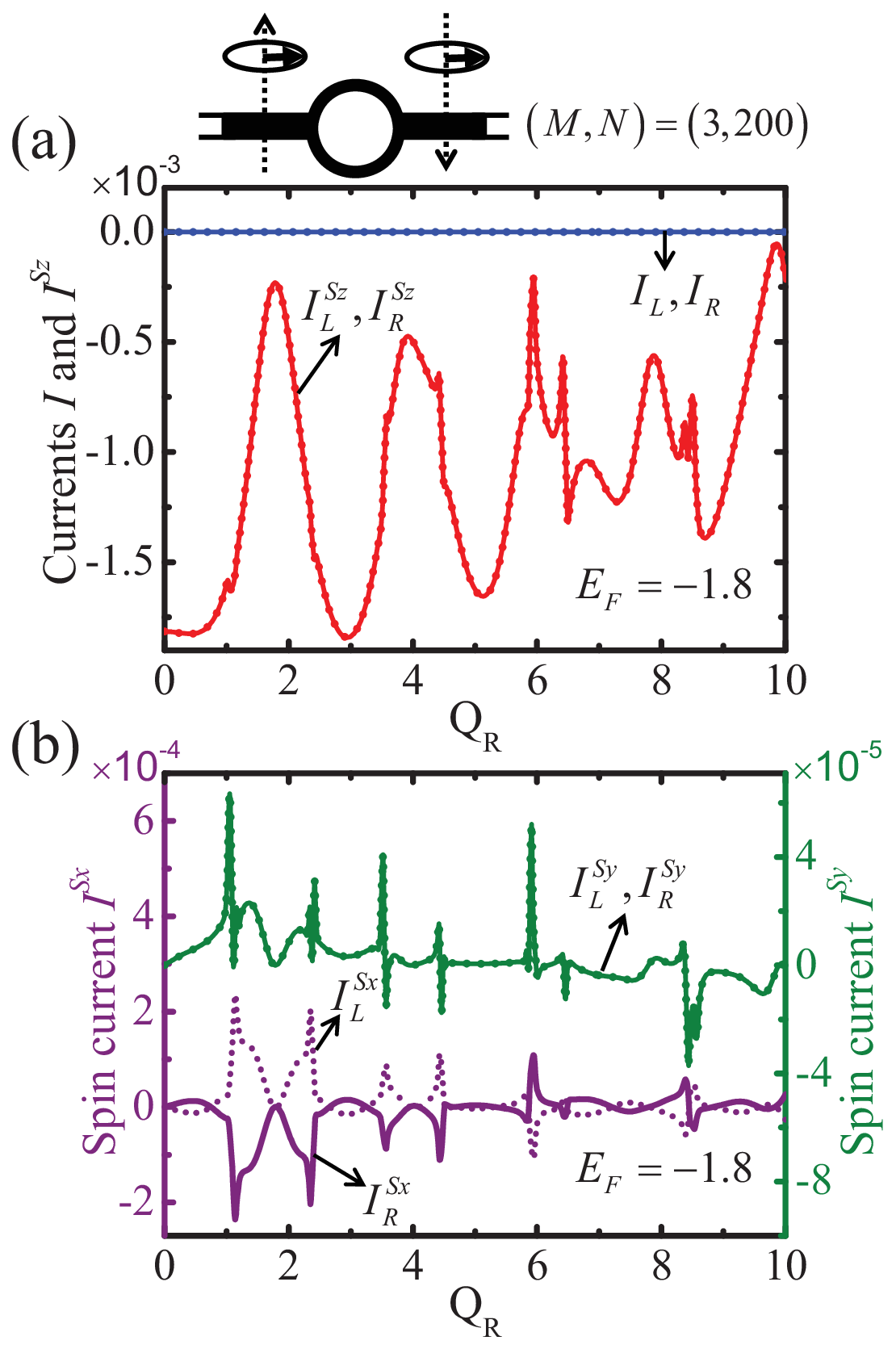,scale=0.45,angle=0,clip=true,trim=0mm 0mm
0mm 0mm}}
\caption{(Color online) Pumped (a) charge and spin-$z$ currents and (b) spin-%
$x$ and spin-$y$ currents versus the dimensionless Rashba spin-orbit
coupling strength $Q_{R}$ in the two-precessing-FM setup $P_{90^{\circ }}$-$%
\bar{P}_{90^{\circ }}$ (top schematics) same as the one considered in Fig.
\protect\ref{fig14:H_LPRP} but with precession axis along $+z$ ($-z$)
direction for the left (right) ferromagnet.}
\label{fig16:H_LPRAP}
\end{figure}

\begin{figure}[tbp]
\centerline{\psfig{file=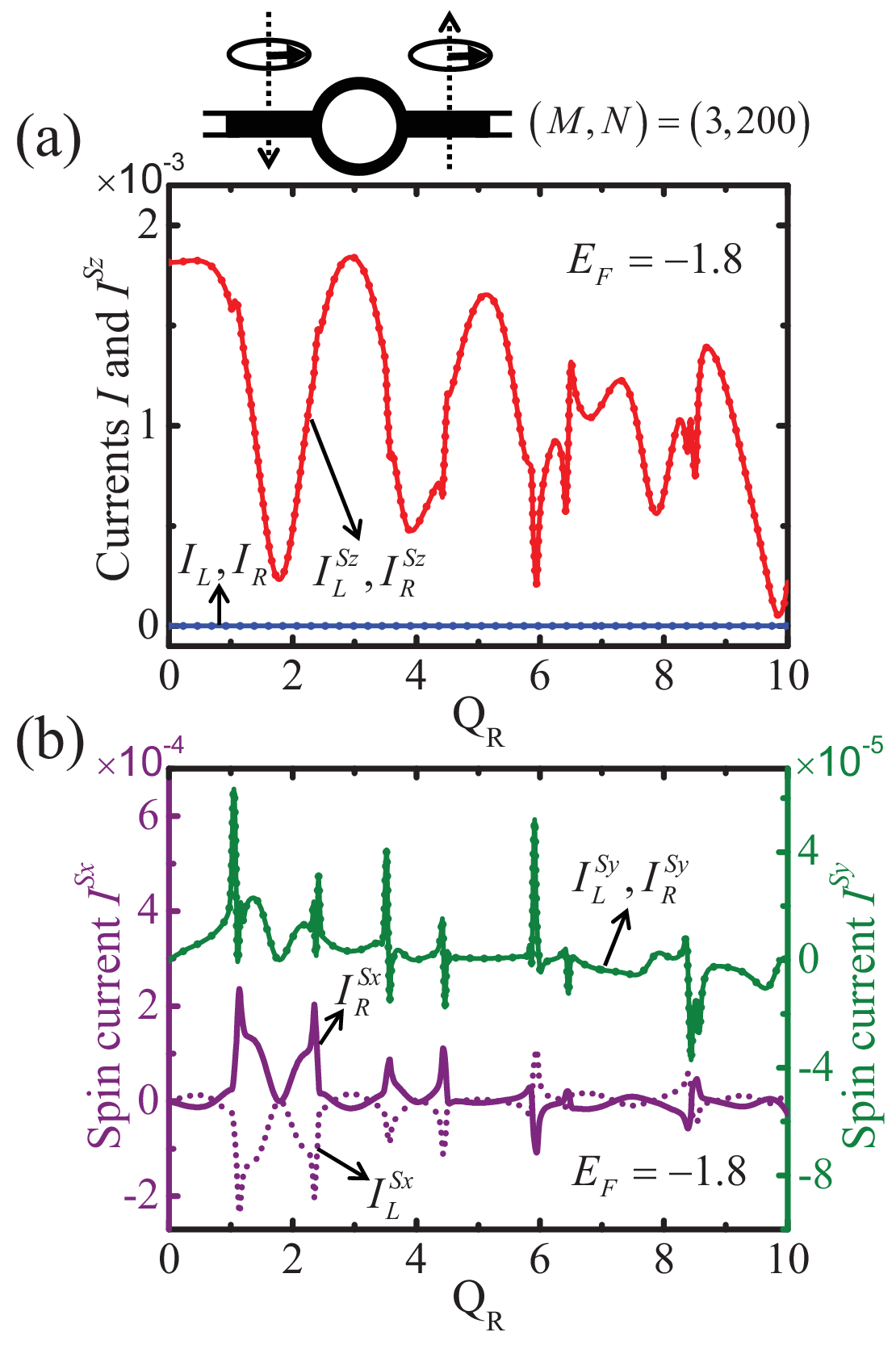,scale=0.45,angle=0,clip=true,trim=0mm 0mm
0mm 0mm}}
\caption{(Color online) Pumped (a) charge and spin-$z$ currents and (b) spin-%
$x$ and spin-$y$ currents versus the dimensionless Rashba spin-orbit
coupling strength $Q_{R}$ in the two-precessing-FM setup $\bar{P}_{90^{\circ
}}$-$P_{90^{\circ }}$ (top schematics) same as the one considered in Fig.
\protect\ref{fig14:H_LPRP} but with precession axis along $-z$ ($+z$)
direction for the left (right) ferromagnet.}
\label{fig17:H_LAPRP}
\end{figure}

Following the same procedure presented above, one can also apply SOB to any $%
A$-$B$ configuration to relate probed currents in a single pumping
configuration (if SOB does not generate another pumping configuration) or to
relate probed currents between different pumping configurations (if SOB
generates another pumping configuration). Here, we choose to focus on the
special case with $\Theta =90^{\circ }$. Unlike the case of $\Theta
=10^{\circ }$ where we have no relations of the pumped currents between the
two different pumping configurations, $P_{10^{\circ }}$-$P_{10^{\circ }}$
and $\bar{P}_{10^{\circ }}$-$\bar{P}_{10^{\circ }}$, at $\Theta =90^{\circ }$
the pumped currents in $P_{90^{\circ }}$-$P_{90^{\circ }}$ can relate to the
pumped currents in $\bar{P}_{90^{\circ }}$-$\bar{P}_{90^{\circ }}$. Applying
SOB to $P_{90^{\circ }}$-$P_{90^{\circ }}$ (Fig. \ref{fig14:H_LPRP}) yields $%
\bar{P}_{90^{\circ }}$-$\bar{P}_{90^{\circ }}$ (Fig. \ref{fig15:H_LAPRAP})
so that $I_{L(R)}$ in $P_{90^{\circ }}$-$P_{90^{\circ }}$ is equal to $%
I_{L(R)}$ in $\bar{P}_{90^{\circ }}$-$\bar{P}_{90^{\circ }}$. This relation,
again, incorporated with the replacement (\ref{eq:SOB_sigma}) leads to the
spin current $(I_{L(R)}^{S_{x}},I_{L(R)}^{S_{y}},I_{L(R)}^{S_{z}})$ in $%
P_{90^{\circ }}$-$P_{90^{\circ }}$ equal to the spin current $%
(-I_{L(R)}^{S_{x}},I_{L(R)}^{S_{y}},-I_{L(R)}^{S_{z}})$ in $\bar{P}%
_{90^{\circ }}$-$\bar{P}_{90^{\circ }}$. The above predictions agree with
our numerical results, Figs. \ref{fig14:H_LPRP} and \ref{fig15:H_LAPRAP}.
Being worth noting, although $P_{10^{\circ }}$-$P_{10^{\circ }}$, $\bar{P}%
_{10^{\circ }}$-$\bar{P}_{10^{\circ }}$, $P_{90^{\circ }}$-$P_{90^{\circ }}$%
, and $\bar{P}_{90^{\circ }}$-$\bar{P}_{90^{\circ }}$ all generate \emph{net}
pure in-plane ($x$-$y$ plane) spin currents, i.e., $%
I_{L}+I_{R}=I_{L}^{S_{z}}+I_{R}^{S_{z}}=0$ as predicted by SOA, the pumped
spin currents at $\Theta =90^{\circ }$ are one to two order larger than
those at $\Theta =10^{\circ }$ (compare Figs. \ref{fig10:LPRP} and \ref%
{fig11:LAPRAP} with Figs. \ref{fig14:H_LPRP} and \ref{fig15:H_LAPRAP}).
Similar enhancement of the pumped spin currents by the cone angle can also
be found by comparing Figs. \ref{fig12:LPRAP} ($P_{10^{\circ }}$-$\bar{P}%
_{10^{\circ }}$) and \ref{fig13:LAPRP} ($\bar{P}_{10^{\circ }}$-$%
P_{10^{\circ }}$) with Figs. \ref{fig16:H_LPRAP} ($P_{90^{\circ }}$-$\bar{P}%
_{90^{\circ }}$) and \ref{fig17:H_LAPRP} ($\bar{P}_{90^{\circ }}$-$%
P_{90^{\circ }}$). The additional (beside what were obtained by SOA)
relations between $P_{90^{\circ }}$-$\bar{P}_{90^{\circ }}$ and $\bar{P}%
_{90^{\circ }}$-$P_{90^{\circ }}$ can be obtained by performing SOB.
Applying SOB to $P_{90^{\circ }}$-$\bar{P}_{90^{\circ }}$ (Fig. \ref%
{fig16:H_LPRAP}), we arrive at $\bar{P}_{90^{\circ }}$-$P_{90^{\circ }}$
(Fig. \ref{fig17:H_LAPRP}), with $I_{L(R)}$ in $P_{90^{\circ }}$-$\bar{P}%
_{90^{\circ }}$ equal to $I_{L(R)}$ in $\bar{P}_{90^{\circ }}$-$P_{90^{\circ
}}$, and then, by replacement (\ref{eq:SOB_sigma}), $%
(I_{L(R)}^{S_{x}},I_{L(R)}^{S_{y}},I_{L(R)}^{S_{z}})$ in $P_{90^{\circ }}$-$%
\bar{P}_{90^{\circ }}$ equal to $%
(-I_{L(R)}^{S_{x}},I_{L(R)}^{S_{y}},-I_{L(R)}^{S_{z}})$ in $\bar{P}%
_{90^{\circ }}$-$P_{90^{\circ }}$, in line with Figs. \ref{fig16:H_LPRAP}
and \ref{fig17:H_LAPRP}. We note that for all $\Theta =90^{\circ }$ pumping
configurations, the pumped spin currents are pure (namely, $I_{L}=I_{R}=0$);
this again can be achieved by considering the current conservation together
with the symmetry argument based on SOA and SOB. We emphasize that our
results here show that the current polarization direction can be tuned by
the top gate voltage governing $Q_{R}$, and the magnitude of the pumped
currents can be controlled by the precession cone angle $\Theta $, offering
amenable manipulations on the output currents from the proposed devise.

\section{Conclusion}

\label{sec:conc}

In conclusion, by introducing the auxiliary system where the time domain is
treated effectively as an additional real-space degree of freedom, we offer
a plain access to the Floquet-NEGF formalism capable of dealing with
time-periodic dynamic problems in full quantum approach. Particularly, by
further adopting the reduced-zone scheme,~\cite{Tsuji2008} we derived
expressions (\ref{eq:I_pt}), (\ref{eq:I_pt^S_q}), (\ref{eq:I_p}), and (\ref%
{eq:I_p^S_q}) in which charge currents are conserved for any given maximum
number of photons. With the help of Eqs. (\ref{eq:I_p}) and (\ref{eq:I_p^S_q}%
), we reveal the physics attributed to the AC effect by considering the ring
of Rashba SOC in spin-driven four- [Fig.~\ref{fig1:setup}(a)] and
two-terminal [Fig.~\ref{fig1:setup}(b)] setups.

When the number of open channel in the leads is one, $M_{\text{open}}=1$, as
a consequence of the AC effect, the complete AC-spin-interference-induced
modulation nodes characterized by $I_{R}=I_{R}^{S_{z}}=0$ at certain Rashba
SOC strengths $Q_{R}^{\ast }$s (where the complete destructive interferences
occur) are found to be independent of the Fermi energy $E_{F}$ in both two-
and four-terminal cases (Figs.~\ref{fig2:2LLPM1},~\ref{fig3:2LLPM3_1}, and~%
\ref{fig5:LPM1}).

In the four-terminal setup, the interference modulation is also
characterized by $I_{B}=I_{T}=0$ at the corresponding two-terminal $%
Q_{R}^{\ast }$s (i.e., Fig.~\ref{fig5:LPM1}, top panel $%
I_{R}=I_{R}^{S_{z}}=0 $ and bottom panel $I_{B}=I_{T}=0$ vanish at the same $%
Q_{R}^{\ast }$s). Increasing the number of open channels by tuning the
Fermi-energy to reach $M_{\text{open}}>1$ regime in quasi 1D (of finite
width) rings destroys the completeness of the modulation, i.e., absence of $%
Q_{R}^{\ast }$ (Fig.~\ref{fig4:2LLPM3_2}). Nevertheless, in the
four-terminal case, we find that the ISHE identified by $I_{B}=I_{T}$ and $%
I_{B}^{S_{z}}+I_{T}^{S_{z}}=0$ (Fig.~\ref{fig5:LPM1}, bottom panel) is
robust against the ring width (Fig.~\ref{fig6:4LLPM}) and weak disorder
(Fig.~\ref{fig7:dis_4LLPM}), and therefore, the proposed device offers a
durable electrical means to measure the pure spin currents pumped by the
precessing FM islands using the inverse quantum-interference-controlled SHE
in the AC rings. The above features based on our spin-driven setups
reciprocally well correspond to the findings in the electric-driven setup,
Refs. \onlinecite{Souma2004} and ~\onlinecite{Souma2005a}, supporting our
derived formalism.

In addition to single-precessing-FM setup (Fig.~\ref{fig1:setup}),
multiple-precessing-FM setup is studied. In the two-terminal
two-precessing-FM setup where the ring is in contact with two (left and
right) precessing FM islands, we find that the currents probed by the left
(right) lead are independent of $\Theta _{R}$ and $\Phi _{R}$ ($\Theta _{L}$
and $\Phi _{L}$) of the right (left) FM under the condition of complete
destructive interferences. In other words, the complete destructive
interference blocks out the relation between the left portion (left FM and
left lead) and the right portion (right FM and right lead) of our device,
while this relation revives when the condition of the destructive
interference is suppressed (Fig.~\ref{fig8:2FM_Lcu}).

We also identified two symmetry operations, SOA and SOB (Fig.~\ref{fig9:SOs}%
), to examine the relations between currents in the same pumping
configurations or different configurations. Performing SOA or SOB on an
arbitrary pumping configuration together with the fact that charge currents
should be conserved, one can first relate the charge currents either in the
same or in different pumping configurations, and then using Eq. (\ref%
{eq:SOA_sigma}) for SOA or Eq. (\ref{eq:SOB_sigma}) for SOB, one can further
obtain the relations between spin currents. We choose to exemplify how the
above procedure works by considering the two-terminal two-precessing-FM
setup with precession cone angles $\Theta _{L}=\Theta _{R}=10^{\circ }$
(Figs.~\ref{fig10:LPRP},~\ref{fig11:LAPRAP},~\ref{fig12:LPRAP}, and~\ref%
{fig13:LAPRP}) and $\Theta _{L}=\Theta _{R}=90^{\circ }$ (Figs.~\ref%
{fig14:H_LPRP},~\ref{fig15:H_LAPRAP},~\ref{fig16:H_LPRAP}, and~\ref%
{fig17:H_LAPRP}). The relations predicted by SOA and SOB consist with our
numerical results. Especially, the net pure in-plane spin currents (for $x$-$%
y$ plane with $I_{L}+I_{R}=I_{L}^{S_{z}}+I_{R}^{S_{z}}=0$ in Figs.~\ref%
{fig10:LPRP},~\ref{fig11:LAPRAP},~\ref{fig14:H_LPRP}, and~\ref%
{fig15:H_LAPRAP}, and for $y$-$z$ plane with $%
I_{L}+I_{R}=I_{L}^{S_{x}}+I_{R}^{S_{x}}=0$ in Figs.~\ref{fig16:H_LPRAP} and~%
\ref{fig17:H_LAPRP}) can be achieved, and for all $\Theta _{L}=\Theta
_{R}=90^{\circ }$ pumping configurations, the pumped spin currents are pure,
namely, $I_{L}=I_{R}=0$. Therefore, with employing the spin-pumping device
proposed here, the pumped currents can be controlled with their magnitudes
and polarization directions tunable via the pumping configurations
(including the precessing cone angle) and the applied top-gate voltage that
varies $Q_{R}$), giving potential applications in spintronics-based industry.

\begin{acknowledgments}
S.-H. C., C.-L. C., and C.-R. C. gratefully acknowledge financial support by
the Republic of China National Science Council Grant No.
NSC98-2112-M-002-012-MY3. F. M. are supported by DOE Grant No.
DE-FG02-07ER46374. One of the authors, S.-H. C., would like to thank
Branislav K. Nikoli\'{c} for his valuable stimulating discussions and
suggestions.
\end{acknowledgments}

\newif\ifabfull\abfulltrue



\begin{thebibliography}
\expandafter\ifx\csname natexlab\endcsname\relax
\fi
\expandafter\ifx\csname bibnamefont\endcsname\relax
\fi
\expandafter\ifx\csname bibfnamefont\endcsname\relax
\fi
\expandafter\ifx\csname citenamefont\endcsname\relax
\fi
\expandafter\ifx\csname url\endcsname\relax
\fi
\expandafter\ifx\csname urlprefix\endcsname\relax
\fi
\providecommand{\bibinfo}[2]{#2} \providecommand{\eprint}[2][]{\url{#2}}

\bibitem[D'yakonov and Perel'(1971)]{D'yakonov1971b} \bibinfo{author}{%
\bibfnamefont{M.~I.} \bibnamefont{D'yakonov}} and
\bibinfo{author}{\bibfnamefont{V.~I.}
  \bibnamefont{Perel'}}, \bibinfo{journal}{Phys. Lett. A} \textbf{%
\bibinfo{volume}{35}}, \bibinfo{pages}{459} (\bibinfo{year}{1971}).

\bibitem[Hirsch(1999)]{Hirsch1999}
\bibinfo{author}{\bibfnamefont{J.~E.}
\bibnamefont{Hirsch}}, \bibinfo{journal}{Phys. Rev. Lett.} \textbf{%
\bibinfo{volume}{83}}, \bibinfo{pages}{1834} (\bibinfo{year}{1999}).

\bibitem[Murakami et~al.(2003)Murakami, Nagaosa, and Zhang]{Murakami2003a} %
\bibinfo{author}{\bibfnamefont{S.}~\bibnamefont{Murakami}}, %
\bibinfo{author}{\bibfnamefont{N.}~\bibnamefont{Nagaosa}}, and %
\bibinfo{author}{\bibfnamefont{S.-C.} \bibnamefont{Zhang}}, %
\bibinfo{journal}{Science} \textbf{\bibinfo{volume}{301}}, %
\bibinfo{pages}{1348} (\bibinfo{year}{2003}).

\bibitem[Sinova et~al.(2004)Sinova, Culcer, Niu, Sinitsyn, Jungwirth, and
MacDonald]{Sinova2004} \bibinfo{author}{\bibfnamefont{J.}~%
\bibnamefont{Sinova}}, \bibinfo{author}{\bibfnamefont{D.}~%
\bibnamefont{Culcer}}, \bibinfo{author}{\bibfnamefont{Q.}~\bibnamefont{Niu}}%
, \bibinfo{author}{\bibfnamefont{N.~A.} \bibnamefont{Sinitsyn}}, %
\bibinfo{author}{\bibfnamefont{T.}~\bibnamefont{Jungwirth}}, and
\bibinfo{author}{\bibfnamefont{A.~H.}
  \bibnamefont{MacDonald}}, \bibinfo{journal}{Phys. Rev. Lett.} \textbf{%
\bibinfo{volume}{92}}, \bibinfo{pages}{126603} (\bibinfo{year}{2004}).

\bibitem[Awschalom and Flatt{\'{e}}(2007)]{Awschalom2007} %
\bibinfo{author}{\bibfnamefont{D.~D.} \bibnamefont{Awschalom}} and
\bibinfo{author}{\bibfnamefont{M.~E.}
  \bibnamefont{Flatt{\'e}}}, \bibinfo{journal}{Nat. Phys.} \textbf{%
\bibinfo{volume}{3}}, \bibinfo{pages}{153} (\bibinfo{year}{2007}).

\bibitem[Hankiewicz et~al.(2005)Hankiewicz, Li, Jungwirth, Niu, Shen, and
Sinova]{Hankiewicz2005}
\bibinfo{author}{\bibfnamefont{E.~M.}
\bibnamefont{Hankiewicz}}, \bibinfo{author}{\bibfnamefont{J.}~%
\bibnamefont{Li}}, \bibinfo{author}{\bibfnamefont{T.}~%
\bibnamefont{Jungwirth}}, \bibinfo{author}{\bibfnamefont{Q.}~%
\bibnamefont{Niu}}, \bibinfo{author}{\bibfnamefont{S.-Q.} \bibnamefont{Shen}}%
, and \bibinfo{author}{\bibfnamefont{J.}~\bibnamefont{Sinova}}, %
\bibinfo{journal}{Phys. Rev. B} \textbf{\bibinfo{volume}{72}}, %
\bibinfo{pages}{155305} (\bibinfo{year}{2005}).

\bibitem[Saitoh et~al.(2006)Saitoh, Ueda, Miyajima, and Tatara]{Saitoh2006} %
\bibinfo{author}{\bibfnamefont{E.}~\bibnamefont{Saitoh}}, %
\bibinfo{author}{\bibfnamefont{M.}~\bibnamefont{Ueda}}, \bibinfo{author}{%
\bibfnamefont{H.}~\bibnamefont{Miyajima}}, and \bibinfo{author}{%
\bibfnamefont{G.}~\bibnamefont{Tatara}},
\bibinfo{journal}{Appl. Phys.
Lett.} \textbf{\bibinfo{volume}{88}}, \bibinfo{pages}{182509} (%
\bibinfo{year}{2006}).

\bibitem[{Mosendz} et~al.(2010){Mosendz}, {Pearson}, {Fradin}, {Bauer}, {%
Bader}, and {Hoffmann}]{Mosendz2010} \bibinfo{author}{\bibfnamefont{O.}~%
\bibnamefont{{Mosendz}}},
\bibinfo{author}{\bibfnamefont{J.~E.}
\bibnamefont{{Pearson}}},
\bibinfo{author}{\bibfnamefont{F.~Y.}
\bibnamefont{{Fradin}}},
\bibinfo{author}{\bibfnamefont{G.~E.~W.}
\bibnamefont{{Bauer}}},
\bibinfo{author}{\bibfnamefont{S.~D.}
\bibnamefont{{Bader}}}, and \bibinfo{author}{\bibfnamefont{A.}~%
\bibnamefont{{Hoffmann}}}, \bibinfo{journal}{Phys. Rev. Lett.} \textbf{%
\bibinfo{volume}{104}}, \bibinfo{pages}{046601} (\bibinfo{year}{2010}).

\bibitem[Valenzuela and Tinkham(2006)]{Valenzuela2006} \bibinfo{author}{%
\bibfnamefont{S.~O.} \bibnamefont{Valenzuela}} and \bibinfo{author}{%
\bibfnamefont{M.}~\bibnamefont{Tinkham}}, \bibinfo{journal}{Nature} \textbf{%
\bibinfo{volume}{442}}, \bibinfo{pages}{176} (\bibinfo{year}{2006}).

\bibitem[Werake et~al.(2011)Werake, Ruzicka, and Zhao]{Werake2011} %
\bibinfo{author}{\bibfnamefont{L.~K.} \bibnamefont{Werake}}, %
\bibinfo{author}{\bibfnamefont{B.~A.} \bibnamefont{Ruzicka}}, and %
\bibinfo{author}{\bibfnamefont{H.}~\bibnamefont{Zhao}}, %
\bibinfo{journal}{Phys. Rev. Lett.} \textbf{\bibinfo{volume}{106}}, %
\bibinfo{pages}{107205} (\bibinfo{year}{2011}).

\bibitem[Tserkovnyak et~al.(2005)Tserkovnyak, Brataas, Bauer, and Halperin]%
{Tserkovnyak2005} \bibinfo{author}{\bibfnamefont{Y.}~%
\bibnamefont{Tserkovnyak}}, \bibinfo{author}{\bibfnamefont{A.}~%
\bibnamefont{Brataas}},
\bibinfo{author}{\bibfnamefont{G.~E.~W.}
\bibnamefont{Bauer}}, and
\bibinfo{author}{\bibfnamefont{B.~I.}
  \bibnamefont{Halperin}}, \bibinfo{journal}{Rev. Mod. Phys.} \textbf{%
\bibinfo{volume}{77}}, \bibinfo{pages}{1375} (\bibinfo{year}{2005}).

\bibitem[Uchida et~al.(2011)Uchida, Ota, Adachi, Xiao, Nonaka, Kajiwara,
Bauer, Maekawa, and Saitoh]{Uchida2011} \bibinfo{author}{\bibfnamefont{K.}~%
\bibnamefont{Uchida}}, \bibinfo{author}{\bibfnamefont{T.}~\bibnamefont{Ota}}%
, \bibinfo{author}{\bibfnamefont{H.}~\bibnamefont{Adachi}}, %
\bibinfo{author}{\bibfnamefont{J.}~\bibnamefont{Xiao}}, \bibinfo{author}{%
\bibfnamefont{T.}~\bibnamefont{Nonaka}}, \bibinfo{author}{\bibfnamefont{Y.}~%
\bibnamefont{Kajiwara}},
\bibinfo{author}{\bibfnamefont{G.~E.~W.}
\bibnamefont{Bauer}}, \bibinfo{author}{\bibfnamefont{S.}~%
\bibnamefont{Maekawa}}, and \bibinfo{author}{\bibfnamefont{E.}~%
\bibnamefont{Saitoh}}, \bibinfo{journal}{J. Appl. Phys.} \textbf{%
\bibinfo{volume}{111}}, \bibinfo{pages}{103903} (\bibinfo{year}{2012}).

\bibitem[Ando et~al.(2011)Ando, Takahashi, Ieda, Kurebayashi, Trypiniotis,
Barnes, Maekawa, and Saitoh]{Ando2011} \bibinfo{author}{\bibfnamefont{K.}~%
\bibnamefont{Ando}}, \bibinfo{author}{\bibfnamefont{S.}~%
\bibnamefont{Takahashi}}, \bibinfo{author}{\bibfnamefont{J.}~%
\bibnamefont{Ieda}}, \bibinfo{author}{\bibfnamefont{H.}~%
\bibnamefont{Kurebayashi}}, \bibinfo{author}{\bibfnamefont{T.}~%
\bibnamefont{Trypiniotis}},
\bibinfo{author}{\bibfnamefont{C.~H.~W.}
\bibnamefont{Barnes}}, \bibinfo{author}{\bibfnamefont{S.}~%
\bibnamefont{Maekawa}}, and \bibinfo{author}{\bibfnamefont{E.}~%
\bibnamefont{Saitoh}}, \bibinfo{journal}{Nature Mater.} \textbf{%
\bibinfo{volume}{10}}, \bibinfo{pages}{655} (\bibinfo{year}{2011}).

\bibitem[Rashba(2000)]{Rashba2000}
\bibinfo{author}{\bibfnamefont{E.~I.}
\bibnamefont{Rashba}}, \bibinfo{journal}{Phys. Rev. B} \textbf{%
\bibinfo{volume}{62}}, \bibinfo{pages}{16267} (\bibinfo{year}{2000}).

\bibitem[{Ohe} et~al.(2008){Ohe}, {Takeuchi}, {Tatara}, and {Kramer}]%
{Ohe2008} \bibinfo{author}{\bibfnamefont{J.}~\bibnamefont{{Ohe}}}, %
\bibinfo{author}{\bibfnamefont{A.}~\bibnamefont{{Takeuchi}}}, %
\bibinfo{author}{\bibfnamefont{G.}~\bibnamefont{{Tatara}}}, and %
\bibinfo{author}{\bibfnamefont{B.}~\bibnamefont{{Kramer}}}, %
\bibinfo{journal}{Physica E} \textbf{\bibinfo{volume}{40}}, %
\bibinfo{pages}{1554} (\bibinfo{year}{2008}).

\bibitem[Takeuchi et~al.(2010)Takeuchi, Hosono, and Tatara]{Takeuchi2010} %
\bibinfo{author}{\bibfnamefont{A.}~\bibnamefont{Takeuchi}}, %
\bibinfo{author}{\bibfnamefont{K.}~\bibnamefont{Hosono}}, and %
\bibinfo{author}{\bibfnamefont{G.}~\bibnamefont{Tatara}}, %
\bibinfo{journal}{Phys. Rev. B} \textbf{\bibinfo{volume}{81}}, %
\bibinfo{pages}{144405} (\bibinfo{year}{2010}).

\bibitem[Silsbee et~al.(1979)Silsbee, Janossy, and Monod]{Silsbee1979} %
\bibinfo{author}{\bibfnamefont{R.~H.} \bibnamefont{Silsbee}}, %
\bibinfo{author}{\bibfnamefont{A.}~\bibnamefont{Janossy}}, and %
\bibinfo{author}{\bibfnamefont{P.}~\bibnamefont{Monod}}, %
\bibinfo{journal}{Phys. Rev. B} \textbf{\bibinfo{volume}{19}}, %
\bibinfo{pages}{4382} (\bibinfo{year}{1979}).

\bibitem[Brouwer(1998)]{Brouwer1998}
\bibinfo{author}{\bibfnamefont{P.~W.}
\bibnamefont{Brouwer}}, \bibinfo{journal}{Phys. Rev. B} \textbf{%
\bibinfo{volume}{58}}, \bibinfo{pages}{R10135} (\bibinfo{year}{1998}).

\bibitem[Aharonov and Casher(1984)]{Aharonov1984} \bibinfo{author}{%
\bibfnamefont{Y.}~\bibnamefont{Aharonov}} and \bibinfo{author}{%
\bibfnamefont{A.}~\bibnamefont{Casher}}, \bibinfo{journal}{Phys. Rev. Lett.}
\textbf{\bibinfo{volume}{53}}, \bibinfo{pages}{319} (\bibinfo{year}{1984}).

\bibitem[{Mathur} and {Stone}(1992)]{Mathur1992} \bibinfo{author}{%
\bibfnamefont{H.}~\bibnamefont{{Mathur}}} and \bibinfo{author}{%
\bibfnamefont{A.~D.} \bibnamefont{{Stone}}},
\bibinfo{journal}{Phys. Rev.
Lett.} \textbf{\bibinfo{volume}{68}}, \bibinfo{pages}{2964} (%
\bibinfo{year}{1992}).

\bibitem[Richter(2012)]{Richter2012} \bibinfo{author}{\bibfnamefont{K.}~%
\bibnamefont{Richter}}, \bibinfo{journal}{Physics} \textbf{%
\bibinfo{volume}{5}}, \bibinfo{pages}{22} (\bibinfo{year}{2012}).

\bibitem[Aharonov and Anandan(1987)]{Aharonov1987} \bibinfo{author}{%
\bibfnamefont{Y.}~\bibnamefont{Aharonov}} and \bibinfo{author}{%
\bibfnamefont{J.}~\bibnamefont{Anandan}},
\bibinfo{journal}{Phys. Rev.
Lett.} \textbf{\bibinfo{volume}{58}}, \bibinfo{pages}{1593} (%
\bibinfo{year}{1987}).

\bibitem[Frustaglia et~al.(2004)Frustaglia, {K\"{o}nig}, and MacDonald]%
{Frustaglia2004} \bibinfo{author}{\bibfnamefont{D.}~\bibnamefont{Frustaglia}}%
, \bibinfo{author}{\bibfnamefont{J.}~\bibnamefont{{K\"{o}nig}}}, and
\bibinfo{author}{\bibfnamefont{A.~H.}
  \bibnamefont{MacDonald}}, \bibinfo{journal}{Phys. Rev. B} \textbf{%
\bibinfo{volume}{70}}, \bibinfo{pages}{045205} (\bibinfo{year}{2004}).

\bibitem[Nagasawa et~al.(2012)Nagasawa, Takagi, Kunihashi, Kohda, and Nitta]%
{Nagasawa2012} \bibinfo{author}{\bibfnamefont{F.}~\bibnamefont{Nagasawa}}, %
\bibinfo{author}{\bibfnamefont{J.}~\bibnamefont{Takagi}}, %
\bibinfo{author}{\bibfnamefont{Y.}~\bibnamefont{Kunihashi}}, %
\bibinfo{author}{\bibfnamefont{M.}~\bibnamefont{Kohda}}, and %
\bibinfo{author}{\bibfnamefont{J.}~\bibnamefont{Nitta}}, %
\bibinfo{journal}{Phys. Rev. Lett.} \textbf{\bibinfo{volume}{108}}, %
\bibinfo{pages}{086801} (\bibinfo{year}{2012}).

\bibitem[Nitta et~al.(1997)Nitta, Akazaki, Takayanagi, and Enoki]{Nitta1997} %
\bibinfo{author}{\bibfnamefont{J.}~\bibnamefont{Nitta}}, \bibinfo{author}{%
\bibfnamefont{T.}~\bibnamefont{Akazaki}}, \bibinfo{author}{%
\bibfnamefont{H.}~\bibnamefont{Takayanagi}}, and \bibinfo{author}{%
\bibfnamefont{T.}~\bibnamefont{Enoki}}, \bibinfo{journal}{Phys. Rev. Lett.}
\textbf{\bibinfo{volume}{78}}, \bibinfo{pages}{1335} (\bibinfo{year}{1997}).

\bibitem[Grundler(2000)]{Grundler2000} \bibinfo{author}{\bibfnamefont{D.}~%
\bibnamefont{Grundler}}, \bibinfo{journal}{Phys. Rev. Lett.} \textbf{%
\bibinfo{volume}{84}}, \bibinfo{pages}{6074} (\bibinfo{year}{2000}).

\bibitem[{K\"{o}nig} et~al.(2006){K\"{o}nig}, Tschetschetkin, Hankiewicz,
Sinova, Hock, Daumer, Schafer, Becker, Buhmann, and Molenkamp]{Konig2006} %
\bibinfo{author}{\bibfnamefont{M.}~\bibnamefont{{K\"onig}}}, %
\bibinfo{author}{\bibfnamefont{A.}~\bibnamefont{Tschetschetkin}}, %
\bibinfo{author}{\bibfnamefont{E.~M.} \bibnamefont{Hankiewicz}}, %
\bibinfo{author}{\bibfnamefont{J.}~\bibnamefont{Sinova}}, %
\bibinfo{author}{\bibfnamefont{V.}~\bibnamefont{Hock}}, \bibinfo{author}{%
\bibfnamefont{V.}~\bibnamefont{Daumer}}, \bibinfo{author}{\bibfnamefont{M.}~%
\bibnamefont{Schafer}},
\bibinfo{author}{\bibfnamefont{C.~R.}
\bibnamefont{Becker}}, \bibinfo{author}{\bibfnamefont{H.}~%
\bibnamefont{Buhmann}}, and
\bibinfo{author}{\bibfnamefont{L.~W.}
\bibnamefont{Molenkamp}}, \bibinfo{journal}{Phys. Rev.
Lett.} \textbf{\bibinfo{volume}{96}}, \bibinfo{pages}{076804} (%
\bibinfo{year}{2006}).

\bibitem[Nitta and Bergsten(2007)]{Nitta2007} \bibinfo{author}{%
\bibfnamefont{J.}~\bibnamefont{Nitta}} and \bibinfo{author}{%
\bibfnamefont{T.}~\bibnamefont{Bergsten}}, \bibinfo{journal}{New J. Phys.}
\textbf{\bibinfo{volume}{9}}, \bibinfo{pages}{341} (\bibinfo{year}{2007}).

\bibitem[Frustaglia and Richter(2004)]{Frustaglia2004a} \bibinfo{author}{%
\bibfnamefont{D.}~\bibnamefont{Frustaglia}} and \bibinfo{author}{%
\bibfnamefont{K.}~\bibnamefont{Richter}}, \bibinfo{journal}{Phys. Rev. B}
\textbf{\bibinfo{volume}{69}}, \bibinfo{pages}{235310} (\bibinfo{year}{2004}%
).

\bibitem[Molnar et~al.(2004)Molnar, Peeters, and Vasilopoulos]{Molnar2004} %
\bibinfo{author}{\bibfnamefont{B.}~\bibnamefont{Molnar}}, %
\bibinfo{author}{\bibfnamefont{F.~M.} \bibnamefont{Peeters}}, and %
\bibinfo{author}{\bibfnamefont{P.}~\bibnamefont{Vasilopoulos}}, %
\bibinfo{journal}{Phys. Rev. B} \textbf{\bibinfo{volume}{69}}, %
\bibinfo{pages}{155335} (\bibinfo{year}{2004}).

\bibitem[{Souma} and {Nikoli{\'{c}}}(2004)]{Souma2004} \bibinfo{author}{%
\bibfnamefont{S.}~\bibnamefont{{Souma}}} and \bibinfo{author}{%
\bibfnamefont{B.~K.} \bibnamefont{{Nikoli{\'c}}}},
\bibinfo{journal}{Phys.
Rev. B} \textbf{\bibinfo{volume}{70}}, \bibinfo{pages}{195346} (%
\bibinfo{year}{2004}).

\bibitem[Souma and {Nikoli\'{c}}(2005)]{Souma2005a} \bibinfo{author}{%
\bibfnamefont{S.}~\bibnamefont{Souma}} and \bibinfo{author}{%
\bibfnamefont{B.~K.} \bibnamefont{{Nikoli\'{c}}}},
\bibinfo{journal}{Phys.
Rev. Lett.} \textbf{\bibinfo{volume}{94}}, \bibinfo{pages}{106602} (%
\bibinfo{year}{2005}).

\bibitem[Tserkovnyak and Brataas(2007)]{Tserkovnyak2007} \bibinfo{author}{%
\bibfnamefont{Y.}~\bibnamefont{Tserkovnyak}} and \bibinfo{author}{%
\bibfnamefont{A.}~\bibnamefont{Brataas}}, \bibinfo{journal}{Phys. Rev. B}
\textbf{\bibinfo{volume}{76}}, \bibinfo{pages}{155326} (\bibinfo{year}{2007}%
).

\bibitem[Borunda et~al.(2008)Borunda, Liu, Kovalev, Liu, Jungwirth, and
Sinova]{Borunda2008}
\bibinfo{author}{\bibfnamefont{M.~F.}
\bibnamefont{Borunda}}, \bibinfo{author}{\bibfnamefont{X.}~\bibnamefont{Liu}}%
, \bibinfo{author}{\bibfnamefont{A.~A.}
\bibnamefont{Kovalev}},
\bibinfo{author}{\bibfnamefont{X.-J.}
\bibnamefont{Liu}}, \bibinfo{author}{\bibfnamefont{T.}~%
\bibnamefont{Jungwirth}}, and \bibinfo{author}{\bibfnamefont{J.}~%
\bibnamefont{Sinova}}, \bibinfo{journal}{Phys. Rev. B} \textbf{%
\bibinfo{volume}{78}}, \bibinfo{pages}{245315} (\bibinfo{year}{2008}).

\bibitem[{Tsuji} et~al.(2008){Tsuji}, {Oka}, and {Aoki}]{Tsuji2008} %
\bibinfo{author}{\bibfnamefont{N.}~\bibnamefont{{Tsuji}}}, %
\bibinfo{author}{\bibfnamefont{T.}~\bibnamefont{{Oka}}}, and %
\bibinfo{author}{\bibfnamefont{H.}~\bibnamefont{{Aoki}}}, %
\bibinfo{journal}{\prb} \textbf{\bibinfo{volume}{78}}, %
\bibinfo{pages}{235124} (\bibinfo{year}{2008}).

\bibitem[{Nikoli\'{c}} and Souma(2005)]{Nikolic2005} \bibinfo{author}{%
\bibfnamefont{B.~K.} \bibnamefont{{Nikoli\'c}}} and \bibinfo{author}{%
\bibfnamefont{S.}~\bibnamefont{Souma}}, \bibinfo{journal}{Phys. Rev. B}
\textbf{\bibinfo{volume}{71}}, \bibinfo{pages}{195328} (\bibinfo{year}{2005}%
).

\bibitem[{Martinez}(2003)]{Martinez2003} \bibinfo{author}{%
\bibfnamefont{D.~F.} \bibnamefont{{Martinez}}},
\bibinfo{journal}{J. Phys.
A: Math. Gen.} \textbf{\bibinfo{volume}{36}}, \bibinfo{pages}{9827} (%
\bibinfo{year}{2003}).

\bibitem[Kitagawa et~al.(2011)Kitagawa, Oka, Brataas, Fu, and Demler]%
{Kitagawa2011} \bibinfo{author}{\bibfnamefont{T.}~\bibnamefont{Kitagawa}}, %
\bibinfo{author}{\bibfnamefont{T.}~\bibnamefont{Oka}}, \bibinfo{author}{%
\bibfnamefont{A.}~\bibnamefont{Brataas}}, \bibinfo{author}{%
\bibfnamefont{L.}~\bibnamefont{Fu}}, and \bibinfo{author}{\bibfnamefont{E.}~%
\bibnamefont{Demler}}, \bibinfo{journal}{Phys. Rev. B} \textbf{%
\bibinfo{volume}{84}}, \bibinfo{pages}{235108} (\bibinfo{year}{2011}).

\bibitem[Wang et~al.(2003)Wang, Edmonds, Campion, Zhao, Neumann, Foxon,
Gallagher, and Main]{Wang2003}
\bibinfo{author}{\bibfnamefont{K.~Y.}
\bibnamefont{Wang}},
\bibinfo{author}{\bibfnamefont{K.~W.}
\bibnamefont{Edmonds}},
\bibinfo{author}{\bibfnamefont{R.~P.}
\bibnamefont{Campion}},
\bibinfo{author}{\bibfnamefont{L.~X.}
\bibnamefont{Zhao}},
\bibinfo{author}{\bibfnamefont{A.~C.}
\bibnamefont{Neumann}},
\bibinfo{author}{\bibfnamefont{C.~T.}
\bibnamefont{Foxon}},
\bibinfo{author}{\bibfnamefont{B.~L.}
\bibnamefont{Gallagher}}, and
\bibinfo{author}{\bibfnamefont{P.~C.}
\bibnamefont{Main}}, in \emph{%
\bibinfo{booktitle}{Physics of semiconductors 2002: Proceedings of
  the 26th International Conference on the Physics of Semiconductors held in
  Edinburgh, UK, 29 July-2 August 2002}}, edited by \bibinfo{editor}{%
\bibfnamefont{A.~R.} \bibnamefont{Long}} and \bibinfo{editor}{%
\bibfnamefont{J.~H.} \bibnamefont{Davies}} (%
\bibinfo{publisher}{IOP
publishing, Bristol}, \bibinfo{year}{2003}), vol. \bibinfo{volume}{171} of
\emph{\bibinfo{series}{Instit. Phys. Confer. Ser.}}, p.~\bibinfo{pages}{58}.

\bibitem[Hattori(2008)]{Hattori2008} \bibinfo{author}{\bibfnamefont{K.}~%
\bibnamefont{Hattori}}, \bibinfo{journal}{J. Phys. Soc. Jpn.} \textbf{%
\bibinfo{volume}{77}}, \bibinfo{pages}{034707} (\bibinfo{year}{2008}).

\bibitem[Chen and Chang(2008)]{Chen2008} \bibinfo{author}{%
\bibfnamefont{S.-H.} \bibnamefont{Chen}} and \bibinfo{author}{%
\bibfnamefont{C.-R.} \bibnamefont{Chang}}, \bibinfo{journal}{Phys. Rev. B}
\textbf{\bibinfo{volume}{77}}, \bibinfo{pages}{045324} (\bibinfo{year}{2008}%
).

\bibitem[Liu et~al.(2011)Liu, Wu, Chen, and Chang]{Liu2011} %
\bibinfo{author}{\bibfnamefont{M.-H.} \bibnamefont{Liu}}, %
\bibinfo{author}{\bibfnamefont{J.-S.} \bibnamefont{Wu}}, \bibinfo{author}{%
\bibfnamefont{S.-H.} \bibnamefont{Chen}}, and \bibinfo{author}{%
\bibfnamefont{C.-R.} \bibnamefont{Chang}}, \bibinfo{journal}{Phys. Rev. B}
\textbf{\bibinfo{volume}{84}}, \bibinfo{pages}{085307} (\bibinfo{year}{2011}%
).

\bibitem[Zhang et~al.(2003)Zhang, Xue, and Xie]{Zhang2003a} %
\bibinfo{author}{\bibfnamefont{P.}~\bibnamefont{Zhang}}, \bibinfo{author}{%
\bibfnamefont{Q.-K.} \bibnamefont{Xue}}, and \bibinfo{author}{%
\bibfnamefont{X.~C.} \bibnamefont{Xie}}, \bibinfo{journal}{Phys. Rev. Lett.}
\textbf{\bibinfo{volume}{91}}, \bibinfo{pages}{196602} (\bibinfo{year}{2003}%
).

\bibitem[Hattori(2007)]{Hattori2007} \bibinfo{author}{\bibfnamefont{K.}~%
\bibnamefont{Hattori}}, \bibinfo{journal}{Phys. Rev. B} \textbf{%
\bibinfo{volume}{75}}, \bibinfo{pages}{205302} (\bibinfo{year}{2007}).

\bibitem[Tserkovnyak et~al.(2008)Tserkovnyak, Moriyama, and Xiao]%
{Tserkovnyak2008} \bibinfo{author}{\bibfnamefont{Y.}~%
\bibnamefont{Tserkovnyak}}, \bibinfo{author}{\bibfnamefont{T.}~%
\bibnamefont{Moriyama}}, and
\bibinfo{author}{\bibfnamefont{J.~Q.}
\bibnamefont{Xiao}}, \bibinfo{journal}{Phys. Rev. B} \textbf{%
\bibinfo{volume}{78}}, \bibinfo{pages}{020401} (\bibinfo{year}{2008}).

\bibitem[Chen et~al.(2009)Chen, Chang, Xiao, and Nikoli\'{c}]{Chen2009} %
\bibinfo{author}{\bibfnamefont{S.-H.} \bibnamefont{Chen}}, %
\bibinfo{author}{\bibfnamefont{C.-R.} \bibnamefont{Chang}}, %
\bibinfo{author}{\bibfnamefont{J.~Q.} \bibnamefont{Xiao}}, and %
\bibinfo{author}{\bibfnamefont{B.~K.} \bibnamefont{Nikoli\'{c}}}, %
\bibinfo{journal}{Phys. Rev. B} \textbf{\bibinfo{volume}{79}}, %
\bibinfo{pages}{054424} (\bibinfo{year}{2009}).

\bibitem[Mahfouzi et~al.(2012)Mahfouzi, Fabian, Nagaosa, and Nikoli\ifmmode~%
\'{c}\else\'{c}\fi{}]{Mahfouzi2012} \bibinfo{author}{\bibfnamefont{F.}~%
\bibnamefont{Mahfouzi}}, \bibinfo{author}{\bibfnamefont{J.}~%
\bibnamefont{Fabian}}, \bibinfo{author}{\bibfnamefont{N.}~%
\bibnamefont{Nagaosa}}, and
\bibinfo{author}{\bibfnamefont{B.~K.}
  \bibnamefont{Nikoli\ifmmode~\acute{c}\else \'{c}\fi{}}}, %
\bibinfo{journal}{Phys. Rev. B} \textbf{\bibinfo{volume}{85}}, %
\bibinfo{pages}{054406} (\bibinfo{year}{2012}).

\bibitem[Moskalets and B\"uttiker(2002)]{Moskalets2002} \bibinfo{author}{%
\bibfnamefont{M.}~\bibnamefont{Moskalets}} and \bibinfo{author}{%
\bibfnamefont{M.}~\bibnamefont{B\"uttiker}}, \bibinfo{journal}{Phys. Rev. B}
\textbf{\bibinfo{volume}{66}}, \bibinfo{pages}{205320} (\bibinfo{year}{2002}%
).

\bibitem[Wu and Cao(2006)]{Wu2006b}
\bibinfo{author}{\bibfnamefont{B.~H.}
\bibnamefont{Wu}} and
\bibinfo{author}{\bibfnamefont{J.~C.}
\bibnamefont{Cao}}, \bibinfo{journal}{Phys. Rev. B} \textbf{%
\bibinfo{volume}{73}}, \bibinfo{pages}{245412} (\bibinfo{year}{2006}).

\bibitem[Arrachea(2005)]{Arrachea2005} \bibinfo{author}{\bibfnamefont{L.}~%
\bibnamefont{Arrachea}}, \bibinfo{journal}{Phys. Rev. B} \textbf{%
\bibinfo{volume}{72}}, \bibinfo{pages}{125349} (\bibinfo{year}{2005}).

\bibitem[Foa~Torres(2005)]{FoaTorres2005} \bibinfo{author}{%
\bibfnamefont{L.~E.~F.} \bibnamefont{Foa~Torres}},
\bibinfo{journal}{Phys.
Rev. B} \textbf{\bibinfo{volume}{72}}, \bibinfo{pages}{245339} (%
\bibinfo{year}{2005}).

\bibitem[{Wu} and {Cao}(2008)]{Wu2008}
\bibinfo{author}{\bibfnamefont{B.~H.}
\bibnamefont{{Wu}}} and
\bibinfo{author}{\bibfnamefont{J.~C.}
\bibnamefont{{Cao}}}, \bibinfo{journal}{J. Phys.: Condens. Matter} \textbf{%
\bibinfo{volume}{20}}, \bibinfo{pages}{085224} (\bibinfo{year}{2008}).

\bibitem[{Wu} and {Cao}(2010)]{Wu2010}
\bibinfo{author}{\bibfnamefont{B.~H.}
\bibnamefont{{Wu}}} and
\bibinfo{author}{\bibfnamefont{J.~C.}
\bibnamefont{{Cao}}}, \bibinfo{journal}{Phys. Rev. B} \textbf{%
\bibinfo{volume}{81}}, \bibinfo{pages}{085327} (\bibinfo{year}{2010}).

\bibitem[Wu and Timm(2010)]{Wu2010a}
\bibinfo{author}{\bibfnamefont{B.~H.}
\bibnamefont{Wu}} and \bibinfo{author}{\bibfnamefont{C.}~\bibnamefont{Timm}}%
, \bibinfo{journal}{Phys. Rev. B} \textbf{\bibinfo{volume}{81}}, %
\bibinfo{pages}{075309} (\bibinfo{year}{2010}).

\bibitem[Floquet(1883)]{Floquet1883} \bibinfo{author}{\bibfnamefont{G.}~%
\bibnamefont{Floquet}}, \bibinfo{journal}{Ann. Sci. Ec. Normale Super.}
\textbf{\bibinfo{volume}{12}}, \bibinfo{pages}{47} (\bibinfo{year}{1883}).

\bibitem[Haug and Jauho(2007)]{Haug2007} \bibinfo{author}{\bibfnamefont{H.}~%
\bibnamefont{Haug}} and
\bibinfo{author}{\bibfnamefont{A.-P.}
\bibnamefont{Jauho}}, \emph{%
\bibinfo{title}{Quantum kinetics in transport and optics of
  semiconductors}} (\bibinfo{publisher}{Springer, Berlin}, %
\bibinfo{year}{2007}).

\bibitem[Rashba(1960)]{Rashba1960}
\bibinfo{author}{\bibfnamefont{E.~I.}
\bibnamefont{Rashba}}, \bibinfo{journal}{Sov. Phys. Solid State} \textbf{%
\bibinfo{volume}{2}}, \bibinfo{pages}{1109} (\bibinfo{year}{1960}).
\end{thebibliography}
\end{document}